# Quantitative Analysis by the Point-Centered Quarter Method

Kevin Mitchell
Department of Mathematics and Computer Science
Hobart and William Smith Colleges
Geneva, NY 14456

## Contents

**Abstract**

This document is an introduction to the use of the point-centered quarter method. It briefly outlines its history, its methodology, and some of the practical issues (and modifications) that inevitably arise with its use in the field. Additionally this paper shows how data collected using point-centered quarter method sampling may be used to determine importance values of different species of trees and describes and derives several methods of estimating plant density and corresponding confidence intervals.

New to this version is a revision of Appendix D which now includes R Shiny Apps to carry out many of these calculations. References to data sets in the previous version that are no longer publicly available have been removed. These have been replaced with examples and exercises which demonstrate the use of the R Shiny Apps and the earlier R scripts.

Two additional comments are in order. First, this document was originally created for student use in an ecology course. In parts, it retains that flavor with instructions for the neophyte, copious examples, and exercises. Second, this document is better suited for reading online where the extensive links throughout it may be accessed.

# 1 Introduction and History

A wide variety of methods have been used to study forest structure parameters such as population density, basal area, and biomass. While these are sometimes estimated using aerial surveys or photographs, most studies involve measurement of these characteristics for individual trees using a number of different sampling methods. These methods fall into two broad categories: plot-based and plotless. Plot-based methods begin with one or more plots (quadrats, belts) of known area in which the characteristics of interest are measured for each plant. In contrast, plotless methods involve measuring distances for a random sample of trees, typically along a transect, and recording the characteristics of interest for this sample. The point-centered quarter method is one such plotless method.

The advantage to using plotless methods rather than standard plot-based techniques is that they tend to be more efficient. Plotless methods are faster, require less equipment, and may require fewer workers. However, the main advantage is speed. The question, then, is whether accuracy is sacrificed in the process.

Stearns (1949) indicated that the point-centered quarter method dates back at least 150 years and was used by surveyors in the mid-nineteenth century making the first surveys of government land. In the late 1940s and early 1950s, several articles appeared that described a variety of plotless methods and compared them to sampling by quadrats. In particular, Cottam et al. (1953) compared the point-centered quarter method to quadrat sampling and derived empirically a formula that could be used to estimate population density from the distance data collected. Since the current paper is intended as an introduction to these methods, it is worth reminding ourselves what the goal of these methods is by recalling part of the introduction to their paper:

> As our knowledge of plant communities increases, greater emphasis is being placed on the methods used to measure the characteristics of these communities. Succeeding decades have shown a trend toward the use of quantitative methods, with purely descriptive methods becoming less common. One reason for the use of quantitative techniques is that the resulting data are not tinged by the subjective bias of the investigator. The results are presumed to represent the vegetation as it actually exists; any other investigator should be able to employ the same methods in the same communities and secure approximately the same data.

Under the assumption that trees are distributed randomly throughout the survey site, Morisita (1954) provided a mathematical proof for the formula that Cottam et al. (1953) had derived empirically for the estimation of population density using the point-centered quarter method. In other words, the point-centered quarter method could, in fact, be used to obtain accurate estimates of population densities with the advantage that the point-centered quarter method data could be collected more quickly than quadrat data. Subsequently, Cottam and Curtis (1956) provided a more detailed comparison of the point-centered quarter method and three other plotless methods (the closest individual, the nearest neighbor, and the random pairs methods). Their conclusion was:

> The quarter method gives the least variable results for distance determinations, provides more data per sampling point, and is the least susceptible to subjective bias. . .

> It is the opinion of the authors that the quarter method is, in most respects, superior to the other distance methods studied, and its use is recommended.

Beasom and Haucke (1975) compared the same four plotless methods and also concluded that point-centered quarter method provides the most accurate estimate of density. In a comparison of a more diverse set of methods Engeman et al. (1994) have a more nuanced opinion of whether the point-centered quarter method is more efficient in the field and more accurate in its density estimates, especially in situations where individuals are not distributed randomly.

In recent years, as the point-centered quarter method has been used more widely, variations have been proposed by Dahdouh-Guebas and Koedam (2006) to address a number of practical problems that arise in the field (multi-stem trees, quarters where no trees are immediately present).

One use of the point-centered quarter method is to determine the **relative importance** of the various tree species in a community. The term "importance" can mean many things depending on the context. An obvious factor influencing the importance of a species to a community is the number of trees present of that species. However, the importance of some number of small trees is not the same as the importance of the same number of large trees. So the size of the trees also plays a role. Further, how the trees are distributed throughout the community also has an effect. A number of trees of the same species clumped together should have a different importance value than the same number of trees distributed more evenly throughout the community.

Measuring importance can aid understanding the successional stages of a forest habitat. At different stages, different species of trees will dominate. Importance values are one objective way of measuring this dominance.

The three factors that we will use to determine the importance value of a species are the density, the size, and the frequency (distribution). Ideally, to estimate these factors, one would take a large sample, measuring, say, all the trees in a $100 \times 100$ meter square (a hectare). This can be extraordinarily time consuming if the trees are very dense. The point-centered quarter method provides a quick way to make such estimates by using a series of measurements along a transect.

## 2 Materials and Methods

The procedure outlined below describes how to carry out point-centered quarter method data collection along a 100 m transect. It can be scaled up or down, as appropriate, for longer or shorter transects. While this analysis can be carried out alone, groups of two or three can make for very efficient data collection. Material requirements include a 50 or 100 meter tape, a shorter 5 or 10 meter tape, a notebook, a calculator, tablet, or smartphone, and a table of random numbers such as this one at NIST, if your device cannot generate them.

1. Generate a list of 15 to 20 random two-digit numbers. If the difference of any two is 4 or less, cross out the second listed number. There should be 10 or more two-digit numbers remaining; if not, generate additional ones. List the first 10 remaining numbers in increasing order. It is important to generate this list before doing any measurements.
2. Lay out a 100 m transect or longer as needed. Multiple transects may be required to survey large areas.
3. The random numbers represent the distances along the transect at which data will be collected. Random numbers are used to eliminate bias. Everyone always wants to measure that BIG tree along the transect, but such trees may not be representative of the community.[1] Points are at least 5 meters apart to ensure that the same trees will not be measured repeatedly. Caution: If trees are particularly sparse, both the length of the transect and the minimum distance between points may need to be increased.
4. The smallest random number determines the first sampling point along the transect. At this and every sampling point, run an imaginary line perpendicular to the transect. This line and the transect divide the world into four quarters, hence the name, point-centered quarter method.

[1] Even Cottam and Curtis (1956) warn us about this tendency: "Repeated sampling of the same stand with different investigators indicates that some individuals have a tendency to place the sampling points so that large or unusual trees occur more commonly than they occur in the stand."

5. Select one of the quarters. In that quarter, locate the tree nearest to the sampling point. For the purposes of this exercise, to be counted as a "tree" it should have a minimum diameter of 4 cm or, equivalently, a minimum circumference of 12.5 cm. Caution: In other situations, different minimum values may be appropriate.

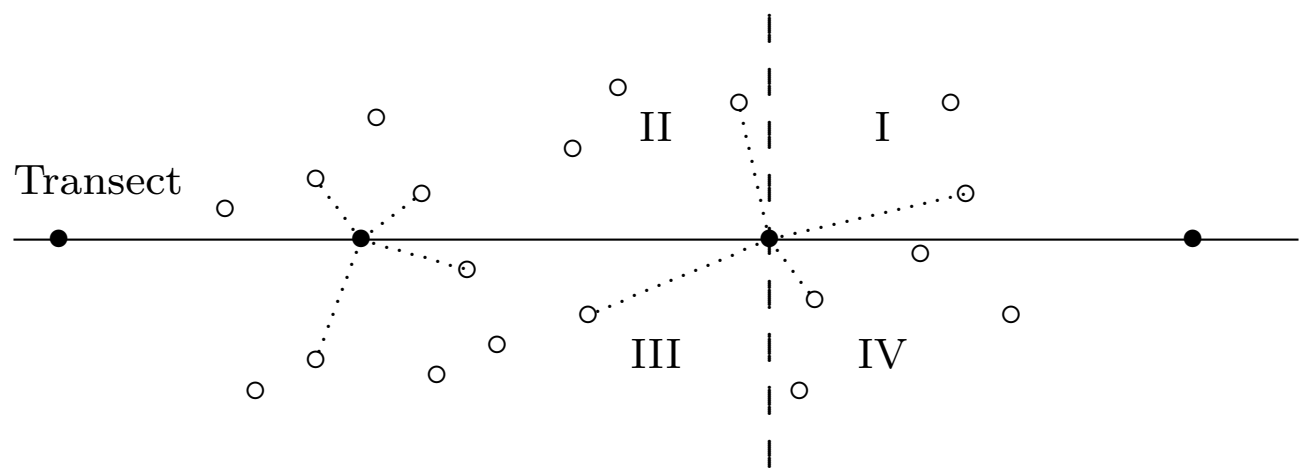

**FIGURE 1.** Sample points along a transect with the nearest trees in each quarter indicated by ······.

For each sampling point, record:

(a) the quarter number (I, II, III, or IV) or other designation (e.g., NE, NW, SE, SW);

(b) the distance from the sampling point to the center of the trunk of the tree, measured to the nearest 0.1 m (Caution: Review Appendix A on the 30–300 Rule.);

(c) the species of the tree;

(d) and the diameter at breast height (DBH) or circumference at chest height (CCH) to the nearest centimeter, but again observe the 30–300 Rule.

**Note:** Brokaw and Thompson (2000) have shown that it is important to use the same height to measure the diameter or circumference. They suggest using a standard height of 130 cm and employing the notation $D_{130}$ rather than DBH to indicate this. Whatever height is used should be explicitly noted in the results.

**Note:** Tree calipers are an easy way to measure diameters, but are often unavailable. It may be more convenient to measure the girth (circumference) of each tree instead of the diameter.

**Cautions:** If a tape is used to measure DBH, avoid protrusions on the trunk. If calipers are used, an average from three caliper readings is recorded. If girths are recorded, rather than convert each girth to a diameter, change the column heading from DBH to CCH. Make the appropriate scaling adjustment in later calculations, whenever diameters are involved.

See Table 1 for how this data should be organized. Repeat this for the other three quarters at this sampling point. If a tree species cannot be identified, simply record it as A, B, C, etc., and collect and label a sample leaf that for comparison purposes at other quarters and later taxonomic identification.

6. Repeat this process for the entire set of sampling points.

7. Carry out the data analysis as described below.

For trees with multiple trunks at breast height, record the diameter (circumference) of each trunk separately. What is the minimum allowed diameter of each trunk in a such multi-trunk tree? Such decisions should be spelled out in the methods section of the resulting report. At a minimum, one should ensure that the combined cross-sectional areas of all trunks meet the previously established minimum cross-sectional area for a single trunk tree. For example, with a 4 cm minimum diameter for a single trunk, the minimum cross-sectional area is

$$\pi r^2 = \pi(2)^2 = 4\pi \approx 12.6 \text{ cm}^2.$$

# 3 Data Organization and Notation

**The Data Layout.** Table 1 illustrates how the data should be organized for the point-centered quarter method analysis. Note the multi-trunk Acacia (8 cm, 6 cm; $D_{130}$) in the third quarter at the second sampling point. The only calculation required at this stage is to sum the distances from the sample points to each of the trees that was measured. **Note**: A sample of only five points as in Table 1 is too few for most studies. These data are presented only to illustrate the method of analysis in a concise way.

**TABLE 1.** Field data organized for point-centered quarter method analysis.

| Sampling Point | Quarter No. | Species | Distance (m) | $D_{130}$ (cm) |
|---|---|---|---|---|
| 1 | 1 | Acacia | 1.1 | 6 |
| | 2 | Eucalyptus | 1.6 | 48 |
| | 3 | Casuarina | 2.3 | 15 |
| | 4 | Callitris | 3.0 | 11 |
| 2 | 1 | Eucalyptus | 2.8 | 65 |
| | 2 | Casuarina | 3.7 | 16 |
| | 3 | Acacia | 0.9 | 8, 6 |
| | 4 | Casuarina | 2.2 | 9 |
| 3 | 1 | Acacia | 2.8 | 4 |
| | 2 | Acacia | 1.1 | 6 |
| | 3 | Acacia | 3.2 | 6 |
| | 4 | Acacia | 1.4 | 5 |
| 4 | 1 | Callitris | 1.3 | 19 |
| | 2 | Casuarina | 0.8 | 22 |
| | 3 | Casuarina | 0.7 | 12 |
| | 4 | Callitris | 3.1 | 7 |
| 5 | 1 | Acacia | 1.5 | 7 |
| | 2 | Acacia | 2.4 | 5 |
| | 3 | Eucalyptus | 3.3 | 27 |
| | 4 | Eucalyptus | 1.7 | 36 |
| | | Total | 40.9 | |

**Notation.** We will use the following notation throughout this paper.

| | |
|---|---|
| $n$ | the number of sample points along the transect |
| $4n$ | the number of samples or observations, one for each quarter at each point |
| $i$ | a particular transect point, where $i = 1, \dots, n$ |
| $j$ | a quarter at a transect point, where $j = 1, \dots, 4$ |
| $R_{ij}$ | the point-to-tree distance at point $i$ in quarter $j$ |

For example, the sum of the distances in the Table 1 is $\sum_{i=1}^{5}\sum_{j=1}^{4} R_{ij} = 40.9$.

# 4 Basic Analysis

The next three subsections outline the estimation of density, frequency, and cover. The most widely studied of the three is density. In Section 5 we present a more robust way to determine both a point estimate and a confidence interval for population density. See Appendix D for apps to quickly carry out all these calculations. In this section density, frequency, and cover are defined both in absolute and relative terms. The relative measures are then combined to create a measure of importance.

## Density

### Absolute Density

The **absolute density** $\lambda$ of trees is defined as the number of trees per unit area. Since $\lambda$ is most easily estimated per square meter and since a hectare is 10,000 m$^2$, $\lambda$ is often multiplied by 10,000 to express the number of tree per hectare. The distances measured using the point-centered quarter method may be used to estimate $\lambda$ to avoid having to count every tree within such a large area.

Note that if $\lambda$ is given as trees/m$^2$, then its reciprocal $1/\lambda$ is the mean area occupied by a single tree. This observation is the basis for the following estimate of $\lambda$. (Also see Section 5.)

From the transect information, determine the **mean distance** $\bar{r}$, which is the sum of the nearest point-to-tree distances in the quarters surveyed divided by the number of quarters,

$$\bar{r} = \frac{\sum_{i=1}^{n} \sum_{j=1}^{4} R_{ij}}{4n}.$$

For the data in Table 1,

$$\bar{r} = \frac{40.9}{20} = 2.05 \text{ m}.$$

Cottam et al. (1953) showed empirically and Morisita (1954) demonstrated mathematically that $\bar{r}$ is actually an estimate of $\sqrt{1/\lambda}$, the square root of the mean area occupied by a single tree. Consequently, an estimate of the density is given by

$$\text{Absolute density} = \hat{\lambda} = \frac{1}{\bar{r}^2} = \frac{16n^2}{\left(\sum_{i=1}^{n} \sum_{j=1}^{4} R_{ij}\right)^2}. \tag{1}$$

For the data in Table 1,

$$\hat{\lambda} = \frac{1}{\bar{r}^2} = \frac{1}{2.05^2} = 0.2380 \text{ trees/m}^2,$$

or, equivalently, 2380 trees/ha.

One way to "see" this is to imagine a "forest" where the trees are uniformly distributed on a square grid whose sides are $\bar{r} = 2.05$ m long. If a tree is located at the center of each square in this forest, then the mean distance $\bar{r}$ between trees is 2.05 m. Such a forest (orchard) is illustrated in Figure 2. Each tree occupies a square of side 2.05 m, and so the density is $1/2.05^2 = 0.2380$ trees/m$^2$. Though such a uniform arrangement of trees violates the assumption of randomness, the figure does illustrate what is happening "on average" or in the mean. (See Appendix B for a careful derivation of this estimate.)

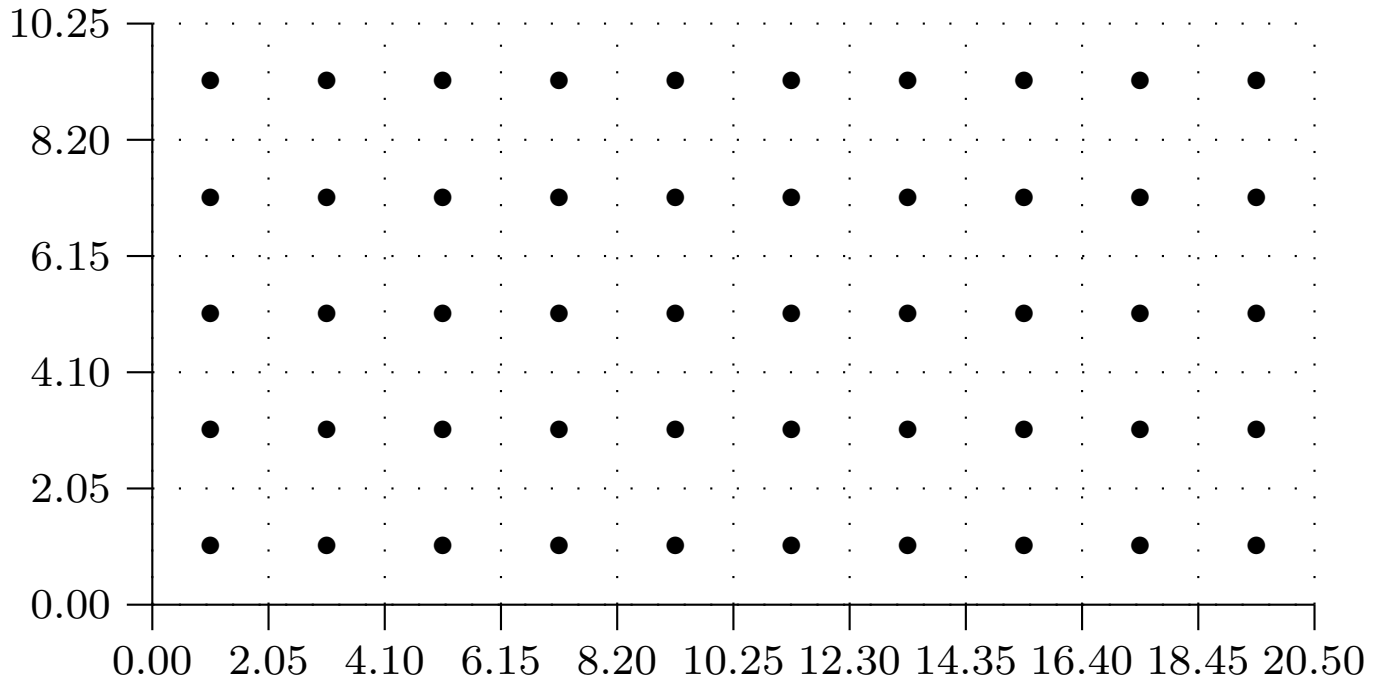

**FIGURE 2.** A grid-like forest with trees uniformly dispersed so that the nearest neighbor is 2.05 m.

### Absolute Density of Each Species

The absolute density of an individual species is the expected number of trees of that species per square meter (or hectare). The **absolute density $\lambda_k$ of species $k$** is estimated as the proportion of quarters in which the species is found times the estimated absolute density of all trees.

$$\hat{\lambda}_k = \frac{\text{Quarters with species } k}{4n} \times \hat{\lambda}. \tag{2}$$

Table 2 gives the absolute density for each species in Table 1.

**TABLE 2.** The absolute density of each species.

| Species | Frequency/Quarter | Trees/ha |
|---|---|---|
| Acacia | $8/20 = 0.40$ | $0.40 \times 2380 = 952$ |
| Eucalyptus | $4/20 = 0.20$ | $0.20 \times 2380 = 476$ |
| Casuarina | $5/20 = 0.25$ | $0.25 \times 2380 = 595$ |
| Callitris | $3/20 = 0.15$ | $0.15 \times 2380 = 357$ |
| Total | | 2380 |

### Relative Density of a Species

The **relative density** of each species is the percentage of the total number observations of that species,

$$\text{Relative density (Species } k) = \frac{\hat{\lambda}_k}{\hat{\lambda}} \times 100.$$

Equivalently by making use of (2), we may define

$$\text{Relative density (Species } k) = \frac{\text{Quarters with species } k}{4n} \times 100. \tag{3}$$

In the current example, using the first definition, the relative density of a species can be found by making use of the data in column 3 of Table 2. For example,

$$\text{Relative density of Eucalyptus} = \frac{476}{2380} \times 100 = 20.0.$$

Using the alternative method in (3) as a check on earlier calculations we see that the relative density is just the proportion in column 2 of Table 2 times 100. For example,

$$\text{Relative density of Eucalyptus} = \frac{4}{20} \times 100 = 20.0.$$

The relative densities should sum to 100 plus or minus a tiny round-off error.

**TABLE 3.** The relative density of each species.

| Species | Relative Density |
|---|---|
| Acacia | 40.0 |
| Eucalyptus | 20.0 |
| Casuarina | 25.0 |
| Callitris | 15.0 |

Based on simulations, Cottam et al. (1953) suggest that about 30 individuals of a particular species must be present in the total sample before confidence can placed in any statements about relative frequency.

## Cover or Dominance of a Species

### Absolute Cover

The cover or dominance of an individual tree is measured by its **basal area** or cross-sectional area. Let $d$, $r$, $c$, and $A$ denote the diameter, radius, circumference, and basal area of a tree, respectively. Since the area of a circle is $A = \pi r^2$, it is also $A = \pi(d/2)^2 = \pi d^2/4$. Since the circumference is $c = 2\pi r$, then the area is also $A = c^2/4\pi$. Either $A = \pi d^2/4$ or $A = c^2/4\pi$ can be used to determine basal area, depending on whether DBH ($D_{130}$) or CCH was recorded in Table 1.

The first step is to compute the basal area for each tree sampled, organizing the data by species. This is the most tedious part of the analysis. A calculator that can handle lists of data or a spreadsheet with its sort function can be very handy at this stages (see Appendix D). For the data in Table 1, the basal area for each tree was obtained using the formula $A = \pi d^2/4$. For trees with multiple trunks, the basal area for each trunk was computed separately and the results summed (see Acacia in Table 4).

**TABLE 4.** The basal area of each tree.

| **Acacia** | | **Eucalyptus** | | **Casuarina** | | **Callitris** | | **Total** |
|---|---|---|---|---|---|---|---|---|
| **$D_{130}$** (cm) | **Area** ($cm^2$) | **$D_{130}$** (cm) | **Area** ($cm^2$) | **$D_{130}$** (cm) | **Area** ($cm^2$) | **$D_{130}$** (cm) | **Area** ($cm^2$) | |
| 6 | 28.3 | 48 | 1809.6 | 15 | 176.7 | 11 | 95.0 | |
| 8, 6 | 78.5 | 65 | 3318.3 | 16 | 201.1 | 19 | 283.5 | |
| 4 | 12.6 | 27 | 572.6 | 9 | 63.6 | 7 | 38.5 | |
| 6 | 28.3 | 36 | 1017.9 | 22 | 380.1 | | | |
| 6 | 28.3 | | | 12 | 113.1 | | | |
| 5 | 19.6 | | | | | | | |
| 7 | 38.5 | | | | | | | |
| 5 | 19.6 | | | | | | | |
| Total BA | 253.7 | | 6718.4 | | 934.6 | | 417.0 | 8323.7 |
| Mean BA | 31.71 | | 1679.60 | | 186.92 | | 139.00 | 416.19 |

Next, determine the total cover or basal area of the trees in the sample by species, and then calculate the mean basal area for each species.[2]
Be careful when computing the means as the number of trees for each species will differ. Remember that each multi-trunk tree counts as a single tree.

The **absolute cover** or **dominance** of each species is expressed as its basal area per hectare. This is obtained by taking the number of trees per species from Table 2 and multiplying by the corresponding mean basal area in Table 4. The units for cover are $m^2$/ha (not $cm^2$/ha), so a conversion factor is required. For Acacia,

$$\text{Absolute Cover (Acacia)} = 31.71 \text{ cm}^2 \times \frac{952}{\text{ha}} \times \frac{1 \text{ m}^2}{10,000 \text{ cm}^2} = 3.0 \frac{\text{m}^2}{\text{ha}}.$$

**TABLE 5.** The total basal area of each species.

| **Species** | **Mean BA** ($cm^2$) | **Number/ha** | **Total BA/ha** ($m^2$/ha) |
|---|---|---|---|
| Acacia | 31.71 | 952 | 3.0 |
| Eucalyptus | 1679.60 | 476 | 79.9 |
| Casuarina | 186.92 | 595 | 11.1 |
| Callitris | 139.00 | 357 | 5.0 |
| Total Cover/ha | | | 99.0 |

Finally, calculate the total cover per hectare by summing the per species covers.

[2]Note: Mean basal area **cannot** be calculated by finding the mean diameter for each species and then using the formula $A = \pi d^2/4$.

### Relative Cover (Relative Dominance) of a Species

The **relative cover** or **relative dominance** [see Cottam and Curtis (1956)] for a particular species is defined to be the absolute cover for that species divided by the total cover times 100 to express the result as a percentage. For example, for Eucalyptus,

$$\text{Relative cover (Eucalyptus)} = \frac{79.9 \text{ m}^2/\text{ha}}{99.0 \text{ m}^2/\text{ha}} \times 100 = 80.7.$$

The relative covers should sum to 100 plus or minus a tiny round-off error. Note that the relative cover can also be calculated directly from the transect information in Table 4.

$$\text{Relative cover (Species } k) = \frac{\text{Total BA of species } k \text{ along transect}}{\text{Total BA of all species along transect}} \times 100. \tag{4}$$

For example,

$$\text{Relative cover (Eucalyptus)} = \frac{6718.4 \text{ cm}^2}{8323.7 \text{ cm}^2} \times 100 = 80.7.$$

**TABLE 6.** The relative cover of each species.

| Species | Relative Cover |
|---|---|
| Acacia | 3.0 |
| Eucalyptus | 80.7 |
| Casuarina | 11.2 |
| Callitris | 5.1 |

## The Frequency of a Species

### Absolute Frequency of a Species

The **absolute frequency** of a species is the percentage of sample points at which a species occurs. Higher absolute frequencies indicate a more uniform distribution of a species while lower values may indicate clustering or clumping. It is defined as

$$\text{Absolute frequency} = \frac{\text{No. of sample points with a species}}{\text{Total number of sample points}} \times 100. \tag{5}$$

For example,

$$\text{Absolute frequency (Acacia)} = \frac{4}{5} \times 100 = 80\%.$$

Note that absolute frequency is based on the number of sample points, not the number of quarters!

**TABLE 7.** The absolute frequency of each species.

| Species | Absolute Frequency |
|---|---|
| Acacia | $(4/5) \times 100 = 80$ |
| Eucalyptus | $(3/5) \times 100 = 60$ |
| Casuarina | $(3/5) \times 100 = 60$ |
| Callitris | $(2/5) \times 100 = 40$ |
| Total | 240 |

Note that the total will sum to more than 100%.

### Relative Frequency of a Species

To normalize for the fact that the absolute frequencies sum to more than 100%, the **relative frequency** is computed. It is defined as

$$\text{Relative frequency} = \frac{\text{Absolute frequency of a species}}{\text{Total frequency of all species}} \times 100. \tag{6}$$

For example,

$$\text{Relative frequency (Acacia)} = \frac{80}{240} \times 100 = 33.3.$$

The relative frequencies should sum to 100 plus or minus a tiny round-off error.

**TABLE 8.** The relative frequency of each species.

| Species | Relative Frequency |
|---|---|
| Acacia | 33.3 |
| Eucalyptus | 25.0 |
| Casuarina | 25.0 |
| Callitris | 16.7 |

What is the difference between relative frequency and relative density? A high relative frequency indicates that the species occurs near relatively many different sampling points, in other words, the species is well-distributed along the transect. A high relative density indicates that the species appears in a relatively large number of quarters. Consequently, if the relative density is high and the relative frequency is low, then the species must appear in lots of quarters but only at a few points, that is, the species appears in clumps. If both are high, the distribution is relatively even and relatively common along the transect. If the relative density is low (appears in few quarters) and the relative frequency is high(er), then the species must be sparsely distributed (few plants, no clumping).

## The Importance Value of a Species

The **importance value** of a species is defined as the sum of the three relative measures:

$$\text{Importance value} = \text{Relative density} + \text{Relative cover} + \text{Relative frequency}. \tag{7}$$

The importance value gives equal weight to the three factors of relative density, cover, and frequency. This means that small trees (that is, with small basal area) can be dominant only if there are enough of them widely distributed across the transect. The importance value can range from 0 to 300.

For the data in Table 1, even though eucalypti are not very common, because of their size they turn out to be the most important species within the community. Because the importance values sum to 300, to make comparisons easier, many researchers choose to report the relative importance, which is just the importance value divided by 3. See Table 9.

**TABLE 9.** The importance value of each species in Table 1.

| Species | Relative Density | Relative Cover | Relative Frequency | Importance | Relative Importance |
|---|---|---|---|---|---|
| Acacia | 40.0 | 3.0 | 33.3 | 76.3 | 25.4 |
| Eucalyptus | 20.0 | 80.7 | 25.0 | 125.7 | 41.9 |
| Casuarina | 25.0 | 11.2 | 25.0 | 61.2 | 20.4 |
| Callitris | 15.0 | 5.1 | 16.7 | 36.8 | 12.3 |

**EXERCISE 4.1.** Look ahead to the data from Mueller-Dombois and Ellenberg (1974) in Example D.4. Calculate the importance value of each species by hand and verify that the results given there are correct.

**Comment.** Each of the measures that make up relative importance may be calculated without knowing the absolute density of the trees at the site [review (3), (4), and (6)]. In fact, any estimate for the absolute density of all species leads to the same relative densities for each species. Consequently, the actual value of density of the plot is not needed to determine relative importance. However, in most studies, absolute density is one of the parameters of greatest interest. Because of this, there have been a number of different methods to estimate absolute density from point-centered quarter method data proposed in the literature. In the next section we explore one of these and others are discussed in Appendix B. Whichever method is used, relative importance is unaffected.

Pollard (1971) showed that the estimate of Cottam and Curtis (1956) of $\lambda$ in (1) is biased.[3] Nonetheless, this estimate appears widely in the literature and, so, has been used here. Another drawback of the estimate in (1) is that no confidence limits are available for it. We now address both of these issues.

# 5 Population Density Reconsidered

Pollard (1971) and Seber (1982) derived an unbiased estimate of the absolute population density using point-centered quarter method data that we now present. It also has the advantage that it can be used to determine confidence intervals for the density.

## Intuition

The discussion that follows is meant to inform our intuition and by no means constitutes a proof of any of the results. That would require a substantially more sophisticated argument (see Appendix B).

Assume that trees are randomly distributed in the survey area. Think of the random points along the transect as representing "virtual trees." The measured distance $R_{ij}$ is a nearest-neighbor distance from a virtual to a real tree and is an estimate of the actual mean nearest-neighbor tree-to-tree distance.

If an actual tree-to-tree distance were $r$ meters, we could draw circles of radius $r/2$ centered at each tree. See Figure 3. Notice that the circles would not overlap and that only one tree would lie in each circle.
The area of each circle is $\pi(r/2)^2 = \pi r^2/4$ m$^2$. Since there is exactly 1 tree per circle and since the circles don't overlap, the density is 1 tree per $\pi r^2/4$ m$^2$, or equivalently,

$$\frac{4}{\pi r^2} \text{ trees/m}^2.$$

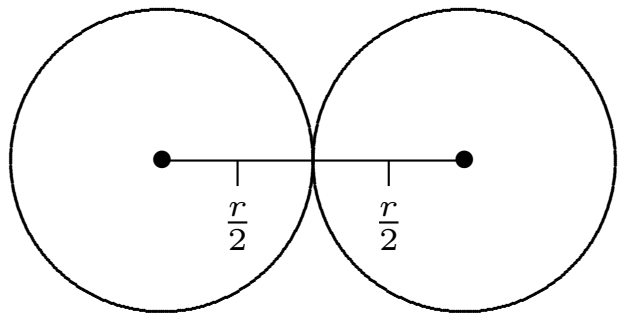

**FIGURE 3.** When trees are $r$ units apart, circles of radius $r/2$ centered at each tree do not overlap and only one tree would lie within each circle.

The observed point-to-tree distances $R_{ij}$ are the estimates of the actual distances. So $\pi(R_{ij}/2)^2 = \pi R_{ij}^2/4$ m$^2$ is an estimate of the sample mean area of a circle occupied by an individual. Using the $4n$ area estimates along the transect, an unbiased estimate of the mean area occupied by an individual is

$$\frac{\sum_{i=1}^{n}\sum_{j=1}^{4}\frac{\pi R_{ij}^2}{4}}{4n-1} = \frac{\pi\sum_{i=1}^{n}\sum_{j=1}^{4}R_{ij}^2}{4(4n-1)}.$$

Note: For this estimate to be unbiased, the denominator is one less than the actual number of observations, that is, $4n-1$. The density is the reciprocal of the mean circular area.
**FORMULA 5.1.** An unbiased estimate of the **population density** $\lambda$ is given by

$$\hat{\lambda} = \frac{4(4n-1)}{\pi\sum_{i=1}^{n}\sum_{j=1}^{4}R_{ij}^2},$$

[3]Pollard (1971) states that the reason for this is Cottam and Curtis (1956) chose to estimate the mean area $A$ occupied by a tree as the reciprocal of $\lambda$. Rather then estimate $A$ directly, as we saw in (1) they estimated $\bar{r}$, which is the reciprocal of the square root of $A$. Squaring and inverting leads to a biased estimate of $A$.

where the units are typically items/m$^2$. Multiplying by 10,000 yields trees/ha. The variance is given by

$$\mathrm{Var}(\hat{\lambda}) = \frac{\hat{\lambda}^2}{4n-2}.$$

**EXAMPLE 5.1.** Reanalyze the data in Table 1 by calculating $\hat{\lambda}$ using Formula 5.1.
**SOLUTION.** First we determine

$$\sum_{i=1}^{n}\sum_{j=1}^{4} R_{ij}^2 = (1.1)^2 + (1.6)^2 + \cdots + (1.7)^2 = 100.71.$$

Unlike in (1), remember to square the distances first, then sum. The density estimate is

$$10,000\hat{\lambda} = 10,000 \cdot \frac{4(4n-1)}{\pi \sum_{i=1}^{n}\sum_{j=1}^{4} R_{ij}^2} = \frac{10,000(4(20-1))}{100.71\pi} = 2402\text{trees/ha}.$$

This estimate is about 1% higher than the earlier biased estimate of 2380.

## Confidence Intervals

For larger samples, approximate confidence interval estimates for $\lambda$ may be calculated in the following way. For details and for an exact confidence interval, see Appendix B. For an R script to compute an exact confidence interval see Appendix D.
**FORMULA 5.2.** For $n > 7$, the endpoints of a confidence interval at the $(1-\alpha)100\%$ level are determined by

$$\text{lower endpoint: } C_1 = \frac{\left(z_{\frac{\alpha}{2}} + \sqrt{16n-1}\right)^2}{\pi \sum_{i=1}^{n}\sum_{j=1}^{4} R_{ij}^2}$$

and

$$\text{upper endpoint: } C_2 = \frac{\left(z_{1-\frac{\alpha}{2}} + \sqrt{16n-1}\right)^2}{\pi \sum_{i=1}^{n}\sum_{j=1}^{4} R_{ij}^2},$$

where $z_\beta$ is the standard normal $z$-score corresponding to probability $\beta$.

**EXAMPLE 5.2.** The data in Table 10 were collected at Lamington National Park in 1994. The data are the nearest point-to-tree distances for each of four quarters at 15 points along a 200 meter transect. The measurements are in meters. Estimate the tree density and find a 95% confidence interval for the density.
**SOLUTION.** In this example, the number of points is $n = 15$ and the number of samples is $4n = 60$. Using Formula 5.1, the point estimate for the density is

$$\hat{\lambda} = \frac{4(4n-1)}{\pi \sum_{i=1}^{n}\sum_{j=1}^{4} R_{ij}^2} = \frac{4(59)}{347.63\pi} = 0.2161 \text{ trees/m}^2.$$

Since the number of points is greater than 7, confidence intervals may be calculated using Formula 5.2. To find a $1-\alpha = 0.95$ confidence interval, we have $\alpha = 0.05$ and so $z_{1-\frac{\alpha}{2}} = z_{0.975} = 1.96$ and $z_{0.025} = -z_{0.975} = -1.96$. The lower endpoint of the confidence interval is

$$C_1 = \frac{\left(z_{0.025} + \sqrt{16n-1}\right)^2}{\pi \sum_{i=1}^{n}\sum_{j=1}^{4} R_{ij}^2} = \frac{\left(-1.96 + \sqrt{16(15)-1}\right)^2}{347.63\pi} = 0.1669$$

**TABLE 10.** Data collected at Lamington National Park in 1994.

| Point | I | II | III | IV |
|---|---|---|---|---|
| 1 | 1.5 | 1.2 | 2.3 | 1.9 |
| 2 | 3.3 | 0.7 | 2.5 | 2.0 |
| 3 | 3.3 | 2.3 | 2.3 | 2.4 |
| 4 | 1.8 | 3.4 | 1.0 | 4.3 |
| 5 | 0.9 | 0.9 | 2.9 | 1.4 |
| 6 | 2.0 | 1.3 | 1.0 | 0.7 |
| 7 | 0.7 | 2.0 | 2.7 | 2.5 |
| 8 | 2.6 | 4.8 | 1.1 | 1.2 |
| 9 | 1.0 | 2.5 | 1.9 | 1.1 |
| 10 | 1.6 | 0.7 | 3.4 | 3.2 |
| 11 | 1.8 | 1.0 | 1.4 | 3.6 |
| 12 | 4.2 | 0.6 | 3.2 | 2.6 |
| 13 | 4.1 | 3.9 | 0.2 | 2.0 |
| 14 | 1.7 | 4.2 | 4.0 | 1.1 |
| 15 | 1.8 | 2.2 | 1.2 | 2.8 |

and the upper endpoint is

$$C_2 = \frac{\left(z_{0.975} + \sqrt{16n-1}\right)^2}{\pi \sum_{i=1}^{n} \sum_{j=1}^{4} R_{ij}^2} = \frac{\left(1.96 + \sqrt{16(15)-1}\right)^2}{347.63\pi} = 0.2778.$$

Therefore, the confidence interval for the density is $(0.1669, 0.2778)$ trees/m$^2$. The units are changed to hectares by multiplying by $10,000$. Thus, $\hat{\lambda} = 2161$ trees/ha while the confidence interval is $(1669, 2778)$ trees/ha.

## Cautions

The estimates and confidence intervals for density assume that the points along the transect are spread out sufficiently so that no individual is sampled in more than one quarter. Further, the density estimate assumes that the spatial distribution of the organisms is completely random. For example, it would be inappropriate to use these methods in an orchard or woodlot where the trees had been planted in rows.

**EXERCISE 5.1.** The following data were collected in interior Alaska by Hollingsworth (2005). The data are the nearest point-to-tree distances in meters for each of four quarters at the first 25 points of 724 sample points. All trees were black spruce, *Picea mariana*. Estimate the tree density and find a 95% confidence interval. [Answer: $\hat{\lambda} = 7037$ trees/ha with a 95% confidence interval of $(5768, 8551)$.]

| Point | I | II | III | IV | Point | I | II | III | IV |
|---|---|---|---|---|---|---|---|---|---|
| 1 | 7.7 | 2.2 | 1.4 | 1.6 | 14 | 1.2 | 1.1 | 1.0 | 1.4 |
| 2 | 0.97 | 1.2 | 1.4 | 1.5 | 15 | 0.5 | 0.7 | 0.9 | 1.1 |
| 3 | 1.4 | 1.4 | 1.8 | 1.6 | 16 | 0.52 | 0.85 | 0.82 | 2.1 |
| 4 | 1.7 | 2.5 | 2.2 | 1.8 | 17 | 0.51 | 0.46 | 1.6 | 1.1 |
| 5 | 0.77 | 1.2 | 1.0 | 1.2 | 18 | 0.46 | 0.9 | 1.7 | 0.65 |
| 6 | 0.38 | 0.64 | 1.84 | 1.7 | 19 | 0.35 | 0.64 | 0.98 | 0.53 |
| 7 | 0.45 | 0.6 | 0.55 | 0.62 | 20 | 0.98 | 1.3 | 2.1 | 1.6 |
| 8 | 0.15 | 0.14 | 0.96 | 0.9 | 21 | 0.35 | 0.5 | 0.25 | 1.0 |
| 9 | 0.39 | 0.5 | 0.57 | 0.88 | 22 | 0.4 | 0.4 | 0.6 | 0.8 |
| 10 | 0.72 | 0.73 | 0.45 | 0.75 | 23 | 0.6 | 1.5 | 1.3 | 1.1 |
| 11 | 0.35 | 1.1 | 0.45 | 1.1 | 24 | 0.4 | 0.5 | 0.9 | 0.8 |
| 12 | 0.55 | 0.9 | 0.65 | 0.9 | 25 | 0.5 | 1.1 | 2.1 | 1.1 |
| 13 | 0.8 | 0.7 | 0.8 | 0.9 | | | | | |

**EXERCISE 5.2.** The following data were collected at Lamington National Park in 1994 by another group of students. The data are the nearest point-to-tree distances (m) for each of four quarters at 14 points along a 200 meter transect. Estimate the tree density and find a 95% confidence interval. [Answer: $\hat{\lambda} = 1382$ trees/ha with a 95% confidence interval of $(1057, 1792)$.]

| I | II | III | IV | I | II | III | IV |
|---|---|---|---|---|---|---|---|
| 0.6 | 1.4 | 3.6 | 2.0 | 3.4 | 3.4 | 2.9 | 2.6 |
| 0.6 | 0.9 | 3.2 | 1.8 | 1.7 | 3.2 | 2.7 | 4.2 |
| 2.0 | 3.9 | 1.8 | 2.2 | 3.8 | 4.2 | 3.2 | 4.4 |
| 4.1 | 7.0 | 1.6 | 4.0 | 1.8 | 1.1 | 4.3 | 3.4 |
| 3.2 | 2.0 | 1.0 | 3.8 | 2.8 | 0.9 | 2.7 | 2.3 |
| 2.8 | 3.3 | 1.3 | 0.8 | 1.4 | 5.0 | 4.5 | 2.7 |
| 3.1 | 1.9 | 2.9 | 3.4 | 2.0 | 0.2 | 3.0 | 4.0 |

# 6 Modifications, Adaptations, and Applications

In Section 1, we indicated that the point-centered quarter method is both efficient and accurate. However, as Díaz et al. (1975) note, in many situations there is

> a discrepancy between the behaviour of the real world and the way it is assumed to behave by the model. Thus, reliability and accuracy have not only a statistical component but also a

> biological one. Most real-life sampling situations violate the assumptions of the underlying models of sampling theory and can render those methods invalid. In such cases, the results may bring about misleading conclusions. In addition, sampling in some environments, such as coastal areas, can be severely constrained by practical considerations.

The material in this section addresses some of these "practical considerations" that occur in the field.

## The Problem with "Breast Height" (BH)

Brokaw and Thompson (2000) did an extensive survey of the literature and found that more than half the papers that used BH did not report the actual value used. Of those that did report BH, values ranged from 120 cm to 160 cm. See Table 11.

**TABLE 11.** The distribution of values stated for "breast height" (BH) in papers published in *Biotropica, Ecology, Journal of Tropical Ecology, Forest Service*, and *Forest Ecology and Management* during the period 1988–1997. Adapted from Brokaw and Thompson (2000), Table 1.

| **BH** (cm) | **120** | **130** | **135** | **137** | **140** | **150** | **160** | **None** | **Total** |
|---|---|---|---|---|---|---|---|---|---|
| Articles | 1 | 113 | 2 | 28 | 27 | 10 | 1 | 258 | 440 |

Since the mode of the BH-values listed was 130 cm, Brokaw and Thompson (2000) strongly suggest adopting this as the standard BH-value. They strongly suggest denoting this value by $D_{130}$ rather than DBH while reserving DBH as a generic term. At a minimum, the BH-value used should be explicitly stated. If a value $x$ other than 130 cm is used, it might be denoted as $D_x$.

As one would expect, DBH does decrease as height increases. In a field survey of 100 trees, Brokaw and Thompson (2000) found that the mean difference between $D_{130}$ and $D_{140}$ was 3.5 mm ($s = 5.8$, $n = 100$). This difference matters. Brokaw and Thompson (2000) report that this resulted in a 2.6% difference in total basal area. When biomass was was calculated using the equation

$$\ln(\text{dry weight}) = -1.966 + 1.242 \ln(\text{DBH}^2)$$

there was a 4.0% difference.

Using different values of BH within a single survey may lead to erroneous results. Additionally, Brokaw and Thompson's (2000) results show that failing to indicate the value of BH may lead to erroneous comparisons of characteristics such as diameter-class distributions, biomass, total basal area, and importance values between studies.

## Vacant Quarters and Truncated Sampling

A question that arises frequently is whether there is a distance limit beyond which one no longer searches for a tree (or other organism of interest) in a particular quarter. The simple answer is, "No." Whenever possible, it is preferable to make sure that every quadrant contains an individual, even if that requires considerable effort. But as a practical matter, a major reason to use the point-centered quarter method is its efficiency, which is at odds with substantial sampling effort. Additionally, in Section 2 we noted that sample points along the transect should be sufficiently far apart so that the same tree is not sampled at two adjacent transect points. Dahdouh-Guebas and Koedam (2006) suggest that it may be preferable to establish a consistent distance limit for the sampling point to the nearest individual rather than to consider the same individual twice. (Note, however, that Cottam and Curtis (1956) explicitly state that they did not use any method to exclude resampling a tree at adjacent transect points and that resampling did, in fact, occur.)

Whether because a distance limit is established for reasons of efficiency (often called truncated sampling) or to prevent resampling, in practice vacant quarters, that is, quadrants containing no tree, may occur. In such cases the calculation of the absolute density must be corrected, since a density calculated from only those quarters containing observations will overestimate the true density.

Warde and Petranka (1981) give a careful derivation of a correction factor (CF) to be used in such cases. In the language of the current paper, as usual, let $n$ denote the number of sampling points and $4n$ the number of quarters. Let $n_0$ denote the number of vacant quarters.

Begin by computing the density for the $4n - n_0$ non-vacant quarters,

$$\bar{r}' = \frac{\sum_{m=1}^{4n-n_0} R_m}{4n - n_0},$$

where $R_m$ is the distance from tree $m$ to its corresponding transect sample point. This is just the analog to (1). Then

$$\text{Corrected Absolute Density} = \hat{\lambda}_c = \frac{\text{CF}}{(\bar{r}')^2}, \qquad (8)$$

where CF is the correction factor from Table 12 corresponding to the proportion of vacant quarters, $\frac{n_0}{4n}$.

**TABLE 12.** Values of the correction factor (CF) to the density estimate based on the formula of Warde and Petranka (1981).

| $n_0/4n$ | CF | $n_0/4n$ | CF | $n_0/4n$ | CF | $n_0/4n$ | CF |
|---|---|---|---|---|---|---|---|
| 0.005 | 0.9818 | 0.080 | 0.8177 | 0.155 | 0.7014 | 0.230 | 0.6050 |
| 0.010 | 0.9667 | 0.085 | 0.8091 | 0.160 | 0.6945 | 0.235 | 0.5991 |
| 0.015 | 0.9530 | 0.090 | 0.8006 | 0.165 | 0.6877 | 0.240 | 0.5932 |
| 0.020 | 0.9401 | 0.095 | 0.7922 | 0.170 | 0.6809 | 0.245 | 0.5874 |
| 0.025 | 0.9279 | 0.100 | 0.7840 | 0.175 | 0.6742 | 0.250 | 0.5816 |
| 0.030 | 0.9163 | 0.105 | 0.7759 | 0.180 | 0.6676 | 0.255 | 0.5759 |
| 0.035 | 0.9051 | 0.110 | 0.7680 | 0.185 | 0.6610 | 0.260 | 0.5702 |
| 0.040 | 0.8943 | 0.115 | 0.7602 | 0.190 | 0.6546 | 0.265 | 0.5645 |
| 0.045 | 0.8838 | 0.120 | 0.7525 | 0.195 | 0.6482 | 0.270 | 0.5590 |
| 0.050 | 0.8737 | 0.125 | 0.7449 | 0.200 | 0.6418 | 0.275 | 0.5534 |
| 0.055 | 0.8638 | 0.130 | 0.7374 | 0.205 | 0.6355 | 0.280 | 0.5479 |
| 0.060 | 0.8542 | 0.135 | 0.7300 | 0.210 | 0.6293 | 0.285 | 0.5425 |
| 0.065 | 0.8447 | 0.140 | 0.7227 | 0.215 | 0.6232 | 0.290 | 0.5370 |
| 0.070 | 0.8355 | 0.145 | 0.7156 | 0.220 | 0.6171 | 0.295 | 0.5317 |
| 0.075 | 0.8265 | 0.150 | 0.7085 | 0.225 | 0.6110 | 0.300 | 0.5263 |

Note that as the proportion of vacant quarters increases, CF decreases and, consequently, so does the estimate of the density (as it should).

**Caution**: Dahdouh-Guebas and Koedam (2006) propose (without mathematical justification) using a correction factor of $\text{CF}' = 1 - \frac{n_0}{4n}$. While this correction factor also lowers the value of the density based on the trees actually measured, this correction differs substantially from that derived by Warde and Petranka (1981). For example, if 5% of the quarters are vacant, then from Table 12 we find CF = 0.873681 while $\text{CF}' = 0.95$.

## The Problem of Unusual Trees or Tree Clusters

**Single Trunk Splitting.** In Section 2 the problem of trees with multiple trunks was briefly considered. What we had in mind there was a tree whose single trunk split into two or more trunks below breast height (130 cm). See Figure 4.

In such a case, there is an unambiguous distance from the point along the transect to the main trunk of the tree. Further, it is natural to obtain the basal area for the tree as the sum of the basal areas for all of the trunks at breast height.

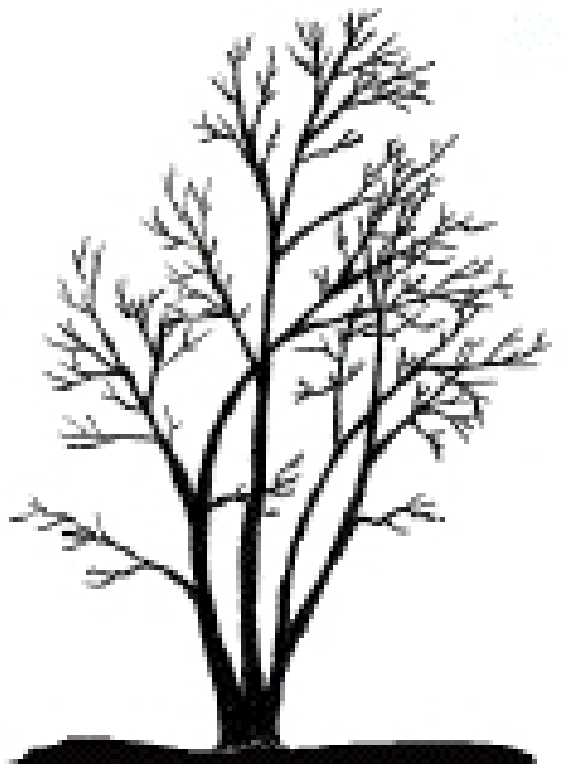

**FIGURE 4.** A tree with a single trunk that splits into multiple trunks below 130 cm.

**Tight Clusters.** However, other configurations of multi-stem trees are possible. A tree may have tightly-clustered multiple trunks at ground level as in Figure 5. In such a case, the entire complex is a single individual. The distance from the transect reference point may be measured to the center of the cluster or, alternatively, be measured as the average of the distances to each of the trunks. As in the previous case, it is natural to obtain the basal area for the tree as the sum of the basal areas for all of the trunks at breast height. (Note: This differs from the the procedure outlined in Dahdouh-Guebas and Koedam (2006) where they suggest using the central stem of the cluster. But they are describing problems with mangroves whose growth architecture is quite different than the trees in the forests of North America. The trees in question here are more similar to those with split trunks.)

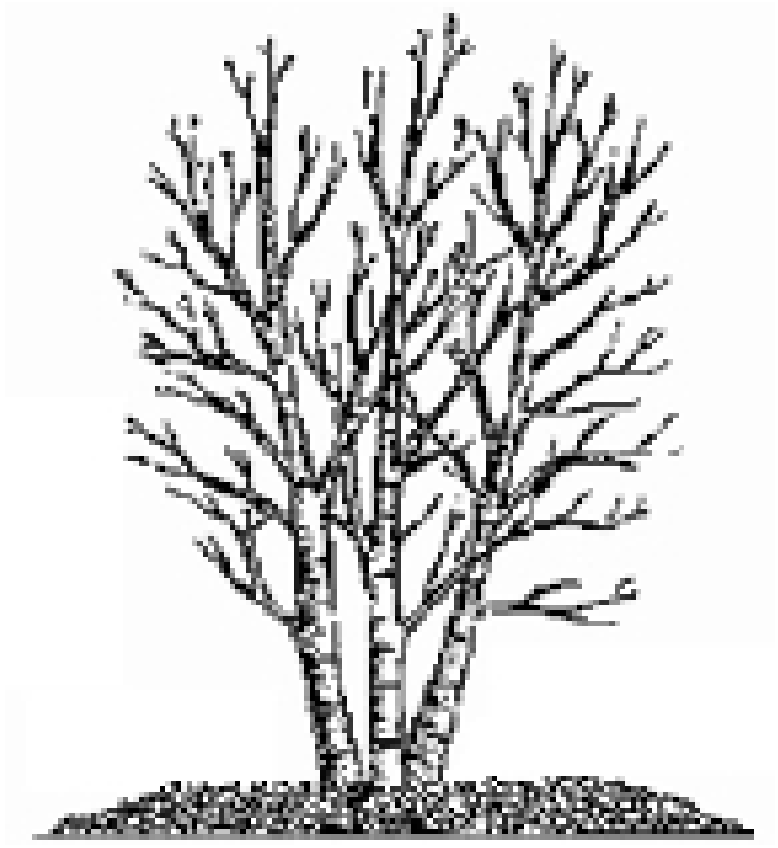

**FIGURE 5.** A birch tree with tightly clustered multiple trunks at ground level.

**Loose Clusters.** Tree clusters such as mangroves present significantly more complicated measurement issues for the point-centered quarter method. Even determining the distance from the transect reference point to such a tree is complicated. Individual stems may be interconnected over relatively large distances, so how does one determine which stems are part of the same individual? The researcher facing such issues is directed to the paper by Dahdouh-Guebas and Koedam (2006) in which they suggest solutions to these and other related questions.

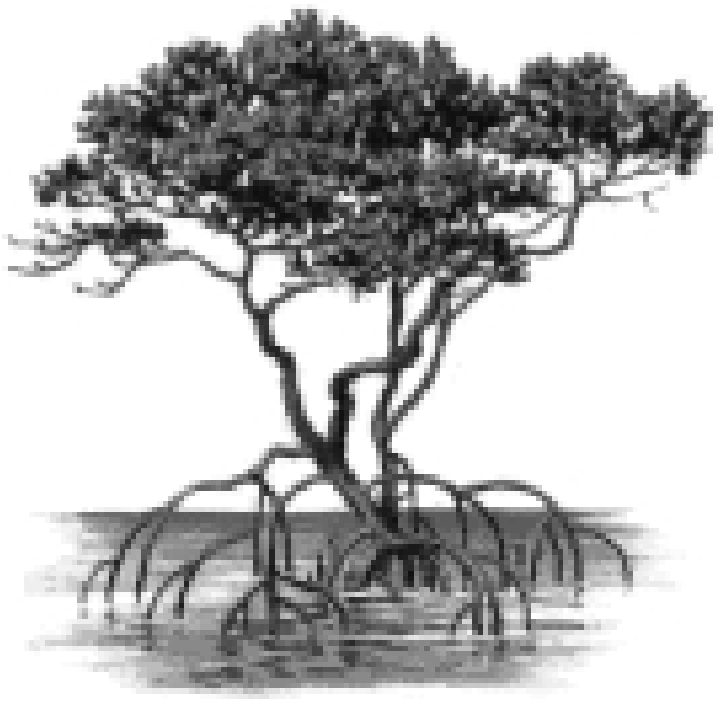

**FIGURE 6.** A individual mangrove with its prop roots has a complex array of roots and stems.

## Miscellaneous Issues

**Crooked Trunks.** In Section 2 we indicated that diameters should be measured at a consistent height and suggested that $D_{130}$ be used. However, some trees may be crooked or growing (somewhat) horizontally at 130 cm above the forest floor. Dahdouh-Guebas and Koedam (2006) suggest that the diameter of such a stem or trunk always be measured at 130 cm *along* the stem, whether or not this is actually 130 cm above the ground.

**Dead Trees.** The implicit but unstated assumption in Section 2 was that we were measuring live trees in the survey. However, depending on the purpose of the survey, dead trees may be important to include. This might be the case if the purpose is to assess exploitable firewood. Such decisions should be explicitly noted in the methods section of the resulting report.

Reversing the roles of live and dead trees, Rheinhardt et al. (1997) used the point-centered quarter method to determine the biomass of standing dead trees in a wetland and also the biomass of coarse woody debris available for nutrient recycling. In the latter case the distance, diameter (minimum 4 in), and length (minimum 3 ft) of the debris item nearest to the transect sampling point in each quarter was recorded.

## Novel Applications

Distance methods have been commonly used for vegetation surveys and are easily adapted to inventories of rare plants or other sessile organisms. The approach may also be useful for population studies of more mobile animal species by obtaining abundance estimates of their nests, dens, roosting sites, or scat piles.

**Nests and Burrows.** Lord Howe Island is a 1455 ha island in the South Pacific Ocean, 495 km east of the Australian mainland. The flesh-footed shearwater (*Puffinus carneipes*) is a medium-sized shearwater with black plumage, pale pinkish feet, and a pale bill with a distinct black tip. On Lord Howe Island these shearwaters nest in single-species colonies at low elevations. The birds excavate burrows up to 3 m long in sandy soils. In 1978, Fullagar and Disney (1981), using the point-centred quarter method and (1), estimated burrow density.

McGrath and Murphy (2012) quantified double-crested cormorant nesting colonies' influence on coarse woody litter and how nest densities and litter depth may influence the herbaceous layer, the seed bank composition and viability across the extent of three Lake Erie islands. Herbaceous layer surveys, cormorant nest counts, soil seed bank cores, and litter depth measurements were all executed using the point-center quarter method to test island-wide impacts from nesting activities.

**Grasslands.** The point-centered quarter method has been adapted to measure density and importance values when sampling grassland vegetation. Dix (1961) used the distance, measured at ground level, from the sampling point to the emergence from the soil of the nearest living herbaceous shoot in each quarter. Since this was the only measurement recorded, importance values were determined using only relative densities and relative frequencies.

Penfound (1963) modified Dix's method to include a relative cover or weight component to better match importance values of trees. In particular, once the distance to a culm or plant was measured, the plant was cut off at soil level and later its oven-dry weight was determined. The relative weight for each species was determined as the total weight for the species divided by the total weight for all species times 100 to express the result as a percentage. The importance of each species was then defined as the sum of the relative frequency, relative density, and relative weight.

On the surface of it, the aggregation often exhibited grassland populations violates the assumption of the random distribution assumption of the point-centered quarter method. Indeed, empirical studies by Risser and Zedler (1968) and Good and Good (1971) indicate that the point-centered quarter method appears to underestimate species density in such cases. In particular, Risser and Zedler (1968) suggest that when using the point-centered quarter method on grasslands, one should check against counts made using quadrat samples.

**Animal Surveys.** The point-centered quarter method was adapted in a series of projects of my students to determine the densities and importance values of certain sessile or relatively slow moving marine organisms.

One group carried out a project surveying holothurians (sea cucumbers) in the reef flat of Lady Elliot Island, a coral cay of the Great Barrier Reef. Transects were laid out in the usual way and the distance and species of the nearest holothurian to each sampling point were recorded for each quarter. These data allowed computation of the relative density and relative frequency for each species.

To take the place of relative cover, the volume of each holothurian was recorded. Volume was estimated by placing each individual in a bucket full of sea water and then removing it. The bucket was then topped off with water from a graduated cylinder and the volume of this water recorded. Since volume and mass are proportional, the relative volume is an approximation of the relative biomass. The sum of the relative density, relative frequency, and relative volume for each species gave its importance value.

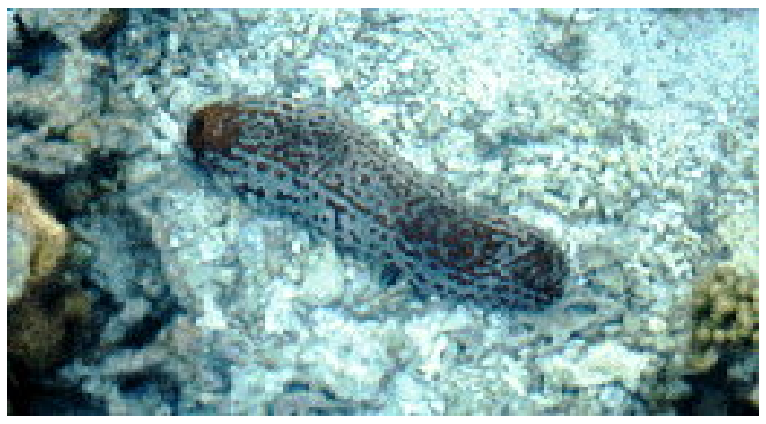

*Bohadschia argus*

A similar survey was conducted both in a reef flat and in an intertidal zone of a sand island for asteroidea (sea stars) using radial "arm length" instead of DBH. Another survey, this time of anemones in the intertidal zone of a sand island was conducted. Since these organisms are more elliptical than circular, major and minor axes were measured from which area occupied could be estimated.

While no extensive testing of the accuracy of these methods was conducted, say against values derived from using quadrats, the use of the point-centered quarter method in each case provided at least a reasonable preliminary snapshot of the relative importance and densities of the organisms surveyed.

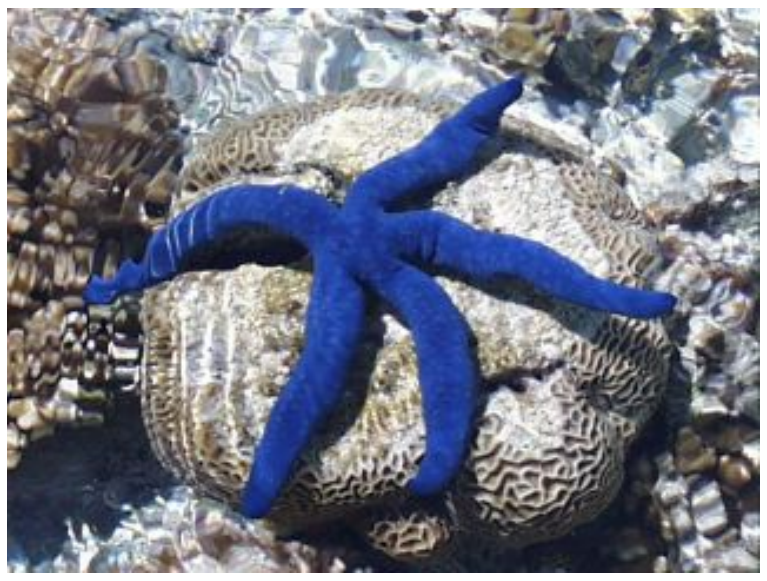

*Linckia laevigata*

**A Final Caution.** Whenever encountering a non-typical situation, it is important to note the situation and its resolution in the resulting report. Be consistent about all such choices. Additional problem issues with possible resolutions are described in Dahdouh-Guebas and Koedam (2006, Appendix B).

# A Accuracy, Precision, and the 30–300 Rule[4]

All biologists are aware of the importance of accuracy and precision in data collection and recording. While these two terms are used synonymously in everyday speech, they have different meanings in statistics. **Accuracy** is the closeness of a measured or computed value to its true value, while **precision** is the closeness of repeated measurements of the same quantity to each other. A biased but sensitive instrument may yield inaccurate but precise readings. On the other hand, an insensitive instrument might result in an accurate reading, but the reading would be imprecise, since another reading of the same object would be unlikely to yield an equally accurate value. Unless there is bias in a measuring instrument, precision will lead to accuracy.

Some measurements are by their nature precise. When we count eggs in a monitor lizard's nest and record the number as 9 or 13, these are exact numbers and, therefore, precise variates. Most continuous variables, however, are approximate with the exact value unknown and unknowable. Recordings of continuous variable data imply a level of precision by the number of digits used. For example, if the length of an adult female monitor lizard is recorded as 97.2 cm, the implied true value of the length is between 97.15 and 97.25 cm. In other words, the last digit recorded defines an interval in which the exact value of the variable resides. A measurement of 97 cm implies a length between 96.5 and 97.5 cm.

In most studies too much precision can slow down data collection while not contributing significantly to the resolution of scientific questions. While it doesn't make sense to measure large eucalyptus trees to the nearest millimeter or to weigh sperm whales to the nearest gram, what level of precision should be recorded? To how many significant figures should we record measurements? Many biologists use the **thirty—three-hundred rule (30–300)** to determine precision for data sets. This rule is easy to apply and will save a great deal of time and effort. Array the sample by order of magnitude from largest to smallest measurement. The number of unit steps between the largest and smallest value should be between 30 and 300. For example, if you were collecting small shells in the intertidal zone of a beach and the largest was 9 mm and the smallest was 5 mm, the number of units steps would be 4 (a unit step is a millimeter in this example). If you recorded the lengths to the nearest tenth of a millimeter with the largest being 9.2 mm and the smallest 5.1 mm in length, the unit step is now 0.1 mm and there are 41 unit steps ($9.2 - 5.1 = 4.1$ mm or 41 tenths of mm) in the data array. The data set will now give you enough precision for most statistical analyses and allow for a reasonable error in recording, that is, a mistake of 1 in the last digit recorded is now less than 2.5% as opposed to 25% when the data were recorded to the nearest millimeter.

[4]The material in this section excerpted from Glover and Mitchell (2008)'s *An Introduction to Biostatics*, Second Edition, page 23.

If sedge plant heights were measured to the nearest tenth of centimeter with the tallest being 194.3 cm and the shortest being 27.1 cm, the unit step would be tenths of centimeters and the data array would have 1672 unit steps ($194.3 - 27.1 = 167.2$ or 1672 tenths of cm). Clearly there is more precision in this data set than is needed. Recording these plant heights to the nearest centimeter would yield 167 unit steps ($194 - 27 = 167$ cm) and would give enough precision for analysis while saving time and effort in data collection.

# B Technical Details

This section outlines the derivation of the density estimate in Section 4 and the estimate and corresponding confidence interval endpoints in Section 5. It also discusses other similar methods of estimating density using plotless methods.

## Derivation of Equation (1)

Assume that a set of points (plants) is distributed randomly over a two-dimensional region where $\lambda$ is the mean number of points per unit area (density). The probability that a randomly chosen region of unit area will contain $x$ points is given by the Poisson distribution

$$\frac{\lambda^x e^{-\lambda}}{x!}.$$

More generally, start with a circle of radius $r$ that is centered at a point chosen at random along a transect. Assume that the circle has been divided into $q$ equiangular sectors and let the region in question be one of these sectors. Then its area is $\pi r^2/q$. If $q = 1$, the region is the entire circle; if $q = 4$ this is the point-centered quarter method. Morisita (1954) used the term "angle methods" to describe density estimates based on this process. The expected number of points in one such sector of the circle is $\lambda\pi r^2/q$ and the so the probability of finding $x$ points in a sector is

$$\frac{(\lambda\pi r^2 q^{-1})^x e^{-\lambda\pi r^2 q^{-1}}}{x!}. \tag{9}$$

Setting $x = 0$, we obtain the probability that a sector of the circle of radius $r$ will contain no points.

$$P(\text{no individuals in a sector circle of radius } r) = e^{-\lambda\pi r^2 q^{-1}}. \tag{10}$$

Equation (10) is a function of $r$ that represents the probability that the distance from the sample point to the nearest individual within the sector is at least $r$. Consequently,

$$P(\text{at least 1 individual in the circle of radius } r) = 1 - e^{-\lambda\pi r^2 q^{-1}}. \tag{11}$$

Differentiating (11) gives the probability density function for $r$

$$f(r) = 2\lambda\pi r q^{-1} e^{-\lambda\pi r^2 q^{-1}}. \tag{12}$$

Therefore, the probability that there is at least one individual in the sector between distances $a$ and $b$ from the center of the circle is

$$\int_a^b 2\lambda\pi r q^{-1} e^{-\lambda\pi r^2 q^{-1}}\, dr. \tag{13}$$

The expected (mean) value of $r$ is obtained by integrating $rf(r)$ over $(0,\infty)$. Using integration by parts and then the substitution $u = \frac{\sqrt{\lambda\pi}}{\sqrt{q}} r$,

$$\begin{aligned}
E(r) &= \int_0^\infty 2\lambda\pi r^2 q^{-1} e^{-\lambda\pi r^2 q^{-1}}\, dr \\
&= -re^{-\lambda\pi r^2 q^{-1}}\Big|_0^\infty + \int_0^\infty e^{-\lambda\pi r^2 q^{-1}}\, dr \\
&= 0 + \frac{\sqrt{q}}{\sqrt{\lambda\pi}} \int_0^\infty e^{-u^2}\, du \\
&= \frac{\sqrt{q}}{\sqrt{\lambda\pi}} \cdot \frac{\sqrt{\pi}}{2} \\
&= \frac{\sqrt{q}}{2\sqrt{\lambda}},
\end{aligned} \tag{14}$$

Solving for the density $\lambda$ in (14) we obtain

$$\lambda = \frac{q}{4[E(r)]^2}. \tag{15}$$

Using the sample mean $\bar{r}$ to estimate $E(r)$ and the point-centered quarter method with $q = 4$, we obtain the estimate of the density in (1),

$$\hat{\lambda} = \frac{1}{\bar{r}^2}$$

As Pollard (1971) and others point out, this estimate is biased.

## Derivation of Formula 5.1

The intuition used in Sections 4 and 5 was that the density and the mean area occupied by a tree are reciprocals of each other. Assume that $n$ random sampling points have been selected along a transect and that there are $q$ equiangular sectors centered at each such point. For $i = 1, \ldots, n$ and $j = 1, \ldots, q$ let $R_{i,j}$ denote the distance from the $i$th sample point to the nearest individual in the $j$th sector. Since these distances are independent, using (12) the likelihood of their joint occurrence is the product

$$\begin{aligned}
&\left(2\lambda\pi R_{1,1} q^{-1} e^{-\lambda\pi R_{1,1}^2 q^{-1}}\right)\left(2\lambda\pi R_{1,2} q^{-1} e^{-\lambda\pi R_{1,2}^2 q^{-1}}\right)\cdots\left(2\lambda\pi R_{n,q} q^{-1} e^{-\lambda\pi R_{n,q}^2 q^{-1}}\right) \\
&\qquad = (2\lambda\pi q^{-1})^{nq}(R_{1,1}R_{12}\cdots R_{n,q})e^{-\lambda\pi q^{-1}\sum_{i=1}^n \sum_{j=1}^q R_{i,j}^2}.
\end{aligned} \tag{16}$$

To simplify notation, denote the $nq$ distances $R_{i,j}$ by $R_m$ for $m = 1, \ldots, nq$ using the one-to-one correspondence $(i,j) \leftrightarrow j+(i-1)q$. For example, $R_{1,1} \leftrightarrow R_1$, $R_{1,q} \leftrightarrow R_q$, $R_{2,1} \leftrightarrow R_{q+1}$, and $R_{n,q} \leftrightarrow R_{nq}$. Then (16) becomes

$$(2\lambda\pi q^{-1})^{nq}(R_1 R_2 \cdots R_{nq})e^{-\lambda\pi q^{-1}\sum_{m=1}^{nq} R_m^2}. \tag{17}$$

Using the $nq$ sample distances, an estimate of the mean area occupied by a tree is given by

$$\frac{\pi q^{-1}\sum_{m=1}^{nq} R_m^2}{nq}.$$

If our intuition is correct, the expected value of the reciprocal of this mean area

$$E\left[\frac{nq}{\pi q^{-1}\sum_{m=1}^{nq} R_m^2}\right] = \int_0^\infty \cdots \int_0^\infty \int_0^\infty \frac{nq}{\pi q^{-1}\sum_{m=1}^{nq} R_m^2}(2\lambda\pi q^{-1})^{nq}(R_1R_2\cdots R_{nq})e^{-\lambda\pi q^{-1}\sum_{m=1}^{nq} R_m^2}\,dR_1dR_2\cdots dR_{nq} \quad (18)$$

should be $\lambda$. To carry out this calculation, use the substitution (see Pollard (1971))

$$u_j = \lambda\pi q^{-1}\sum_{m=1}^{j} R_m^2 \qquad j = 1,\ldots,nq$$

with Jacobian

$$J(u_1,u_2,\ldots,u_{nq}) = \begin{vmatrix} 2\lambda\pi q^{-1}R_1 & 0 & \cdots & 0 \\ 2\lambda\pi q^{-1}R_1 & 2\lambda\pi q^{-1}R_2 & \cdots & 0 \\ \vdots & \vdots & \vdots & \vdots \\ 2\lambda\pi q^{-1}R_1 & 2\lambda\pi q^{-1}R_2 & \cdots & 2\lambda\pi q^{-1}R_{nq} \end{vmatrix} = (2\lambda\pi q^{-1})^{nq}R_1R_2\cdots R_{nq}.$$

The limits of integration for $u_{nq}$ are 0 to $\infty$ and for $u_m$ ($i = m,\ldots,nq-1$) they are 0 to $u_{m+1}$. So (18) becomes

$$\begin{aligned} E\left[\frac{nq}{\pi q^{-1}\sum_{m=1}^{nq} R_m^2}\right] = E\left[\frac{\lambda nq}{u_{nq}}\right] &= \int_0^\infty \cdots \int_0^{u_3}\int_0^{u_2} \frac{\lambda nq}{u_{nq}} e^{-u_{nq}}\,du_1du_2\cdots du_{nq} \\ &= \int_0^\infty \cdots \int_0^{u_3} \frac{\lambda nq u_2}{1\cdot u_{nq}} e^{-u_{nq}}\,du_2\cdots du_{nq} \\ &= \int_0^\infty \cdots \int_0^{u_4} \frac{\lambda nq u_3^2}{2\cdot 1\cdot u_{nq}} e^{-u_{nq}}\,du_3\cdots du_{nq} \\ &\vdots \\ &= \int_0^\infty \frac{\lambda nq u_{nq}^{nq-1}}{(nq-1)!u_{nq}} e^{-u_{nq}}\,du_{nq} \\ &= \frac{\lambda nq}{(nq-1)!}\int_0^\infty u_{nq}^{nq-2}e^{-u_{nq}}\,du_{nq} \\ &= \frac{\lambda nq}{nq-1}. \end{aligned} \quad (19)$$

So the reciprocal of the mean area occupied by a tree is also a biased estimate of $\lambda$, but the bias is easily corrected. An unbiased estimate of the density is

$$\hat\lambda = \frac{nq-1}{nq}\cdot\frac{nq}{\pi q^{-1}\sum_{m=1}^{nq} R_m^2} = \frac{q(nq-1)}{\pi\sum_{i=1}^{n}\sum_{j=1}^{q} R_{ij}^2}. \quad (20)$$

For the point-centered quarter method method where $q = 4$ we have that an unbiased estimate of the density is

$$\hat\lambda = \frac{4(4n-1)}{\pi\sum_{i=1}^{n}\sum_{j=1}^{4} R_{ij}^2},$$

which is Formula 5.1.

It is worth mentioning the interpretation of (20) when $q = 1$. In this case the distance from each sample point to the nearest individual is measured and an unbiased estimate of the density is given by the simpler formula

$$\hat\lambda = \frac{n-1}{\pi\sum_{i=1}^{n} R_i^2}. \quad (21)$$

## Confidence Intervals and the Derivation of Formula 5.2

Next, recall that the probability density function of the chi-square distribution for $x \geq 0$ is

$$f(x;k) = \frac{\left(\frac{1}{2}\right)^{k/2} x^{k/2-1}}{\Gamma(k/2)} e^{-x/2}, \tag{22}$$

where $k$ denotes degrees of freedom and $\Gamma(z)$ is the gamma function.[5] If we let $y = 2\lambda\pi r^2 q^{-1}$, then $dy = 4\lambda\pi r q^{-1}\,dr$ so (13) may be written as

$$\int_{\pi a^2}^{\pi b^2} \tfrac{1}{2} e^{-y/2}\,dy.$$

In other words, using (22) with $k = 2$ and (13) we see that $2\lambda\pi r^2 q^{-1}$ is distributed as $\chi^2_{(2)}$.

To generalize, assume as before that we have selected $n$ random sampling points along a transect and that there are $q$ equiangular sectors centered at each such point. For $i = 1, \ldots, n$ and $j = 1, \ldots, q$ let $R_{ij}$ denote the distance from the $i$th sample point to the nearest individual in the $j$ sector. From (16) the probability of their joint occurrence is the product

$$(2\lambda\pi q^{-1})^{nq}(R_{11}R_{12}\cdots R_{nq})e^{-\lambda\pi q^{-1}\sum_{i=1}^{n}\sum_{j=1}^{q} R_{ij}^2}.$$

Since the distances are independent and since each $2\lambda\pi R_{ij}^2 q^{-1}$ is distributed as $\chi^2_{(2)}$, then

$$2\lambda\pi q^{-1}\sum_{i=1}^{n}\sum_{j=1}^{q} R_{ij}^2 \sim \chi^2_{(2nq)}. \tag{23}$$

Consequently, a $(1-\alpha)100\%$ confidence interval for $\lambda$ is determined by the inequalities

$$\chi_{\frac{\alpha}{2}(2nq)} < 2\lambda\pi q^{-1}\sum_{i=1}^{n}\sum_{j=1}^{q} R_{ij}^2 < \chi_{1-\frac{\alpha}{2}(2nq)}.$$

Solving for $\lambda$ we obtain the following result.

**FORMULA B.1.** Assume $n$ random sampling points have been selected along a transect and that there are $q$ equiangular sectors centered at each such point. For $i = 1, \ldots, n$ and $j = 1, \ldots, q$ let $R_{ij}$ denote the distance from the $i$th sample point to the nearest individual in the $j$th sector. A $(1-\alpha)100\%$ confidence interval for the density $\lambda$ is given by $(C_1, C_2)$, where

$$C_1 = \frac{q\chi_{\frac{\alpha}{2}(2nq)}}{2\pi\sum_{i=1}^{n}\sum_{j=1}^{q} R_{ij}^2} \qquad \text{and} \qquad C_2 = \frac{q\chi_{1-\frac{\alpha}{2}(2nq)}}{2\pi\sum_{i=1}^{n}\sum_{j=1}^{q} R_{ij}^2}.$$

In particular, for the point-centered quarter method where $q = 4$, we have

$$C_1 = \frac{2\chi_{\frac{\alpha}{2}(8n)}}{\pi\sum_{i=1}^{n}\sum_{j=1}^{4} R_{ij}^2} \qquad \text{and} \qquad C_2 = \frac{2\chi_{1-\frac{\alpha}{2}(8n)}}{\pi\sum_{i=1}^{n}\sum_{j=1}^{4} R_{ij}^2}.$$

**EXAMPLE B.1.** Return to Example 5.2 and calculate a confidence interval for the density using Formula B.1.

**SOLUTION.** From Formula B.1,

$$C_1 = \frac{2\chi_{\frac{\alpha}{2}(8n)}}{\pi\sum_{i=1}^{n}\sum_{j=1}^{4} R_{ij}^2} = \frac{2\chi_{\frac{\alpha}{2}(120)}}{\pi\sum_{i=1}^{15}\sum_{j=1}^{4} R_{ij}^2} = \frac{183.15}{1092.11} = 0.1677.$$

and

$$C_2 = \frac{2\chi_{1-\frac{\alpha}{2}(8n)}}{\pi\sum_{i=1}^{n}\sum_{j=1}^{4} R_{ij}^2} = \frac{2\chi_{1-\frac{\alpha}{2}(120)}}{\pi\sum_{i=1}^{15}\sum_{j=1}^{4} R_{ij}^2} = \frac{304.42}{1092.11} = 0.2787.$$

This interval is nearly identical to the one computed in Example 5.2 using a normal approximation.

[5] In particular, if $z$ is a positive integer, then $\Gamma(z) = (z-1)!$.

### Normal Approximation

A difficulty with calculating confidence intervals using Formula B.1 is that $2nq$ is often greater than the degrees of freedom listed in a typical $\chi^2$-table. For larger values of $2nq$, the appropriate $\chi^2$ values can be obtained from a spreadsheet program or other statistical or mathematical software.

Alternatively, one can use a normal approximation. It is a well-known result due to Fisher that if $X \sim \chi^2_{(k)}$, then $\sqrt{2X}$ is approximately normally distributed with mean $\sqrt{2k-1}$ and unit variance. In other words, $\sqrt{2X} - \sqrt{2k-1}$ has approximately a standard normal distribution.

In the case at hand, $2\lambda\pi q^{-1} \sum_{i=1}^{n} \sum_{j=1}^{q} R_{ij}^2 \sim \chi^2_{(2nq)}$. Therefore, the endpoints for a a $(1-\alpha)100\%$ confidence interval for $\lambda$ are determined as follows:

$$\begin{aligned}
&z_{\alpha/2} < \sqrt{2\left(2\lambda\pi q^{-1} \sum_{i=1}^{n} \sum_{j=1}^{q} R_{ij}^2\right)} - \sqrt{2(2nq)-1} < z_{1-\alpha/2} \\
&\iff z_{\alpha/2} + \sqrt{4nq-1} < \sqrt{4\lambda\pi q^{-1} \sum_{i=1}^{n} \sum_{j=1}^{q} R_{ij}^2} < z_{1-\alpha/2} + \sqrt{4nq-1} \\
&\iff \frac{z_{\alpha/2} + \sqrt{4nq-1}}{\sqrt{4\pi q^{-1} \sum_{i=1}^{n} \sum_{j=1}^{q} R_{ij}^2}} < \sqrt{\lambda} < \frac{z_{1-\alpha/2} + \sqrt{4nq-1}}{\sqrt{4\pi q^{-1} \sum_{i=1}^{n} \sum_{j=1}^{q} R_{ij}^2}}.
\end{aligned}$$

Squaring, we find:

**FORMULA B.2.** For $nq > 30$, the endpoints of a $(1-\alpha)100\%$ confidence interval for the density $\lambda$ are well-approximated by

$$C_1 = \frac{\left(z_{\frac{\alpha}{2}} + \sqrt{4nq-1}\right)^2}{4\pi q^{-1} \sum_{i=1}^{n} \sum_{j=1}^{q} R_{ij}^2} \quad \text{and} \quad C_2 = \frac{\left(z_{1-\frac{\alpha}{2}} + \sqrt{4nq-1}\right)^2}{4\pi q^{-1} \sum_{i=1}^{n} \sum_{j=1}^{q} R_{ij}^2}.$$

For the point-centered quarter method where $q = 4$ we obtain

$$C_1 = \frac{\left(z_{\frac{\alpha}{2}} + \sqrt{16n-1}\right)^2}{\pi \sum_{i=1}^{n} \sum_{j=1}^{4} R_{ij}^2} \quad \text{and} \quad C_2 = \frac{\left(z_{1-\frac{\alpha}{2}} + \sqrt{16n-1}\right)^2}{\pi \sum_{i=1}^{n} \sum_{j=1}^{4} R_{ij}^2}.$$

Note that the later formula above is Formula 5.2.

## Further Generalizations: Order Methods

Order methods describe the estimation of the density $\lambda$ by measuring the distances from the sample point to the first, second, third, etc. closest individuals. Note: The data collected during point-centered quarter method sampling (as in Table 1) do *not* necessarily measure the first through fourth closest individuals to the sample point because any two, three, or four closest individuals may lie in a single quadrant or at least be spread among fewer than all four quadrants.

The derivation that follows is an adaptation of Moore (1954), Seber (1982), Eberhardt (1967), and Morisita (1954). We continue to assume, as above, that the population is randomly distributed with density $\lambda$ so that the number of individuals $x$ in a circle of radius $r$ chosen at random has a Poisson distribution

$$P(x) = \frac{(\lambda\pi r^2)^x e^{-\lambda\pi r^2}}{x!}.$$

Let $R_{(k)}$ denote the distance to the $k$th nearest tree from a random sampling point. Then

$$\begin{aligned}
P(R_{(k)} \le r) &= P(\text{finding at least } k \text{ individuals in a circle of area } \pi r^2) \\
&= \sum_{i=k}^{\infty} e^{-\lambda\pi r^2} \left[\frac{\left(\lambda\pi r^2\right)^i}{i!}\right]. \qquad (24)
\end{aligned}$$

Taking the derivative of (24), the corresponding pdf for $r$ is

$$\begin{aligned} f_k(r) &= \sum_{i=k}^{\infty}\left(-2\lambda\pi r e^{-\lambda\pi r^2}\left[\frac{\left(\lambda\pi r^2\right)^i}{i!}\right] + e^{-\lambda\pi r^2}\left[\frac{2i\lambda\pi r\left(\lambda\pi r^2\right)^{(i-1)}}{i!}\right]\right) \\ &= 2\lambda\pi r e^{-\lambda\pi r^2}\sum_{i=k}^{\infty}\left(-\frac{\left(\lambda\pi r^2\right)^i}{i!} + \frac{\left(\lambda\pi r^2\right)^{(i-1)}}{(i-1)!}\right) \\ &= \frac{2\lambda\pi r e^{-\lambda\pi r^2}\left(\lambda\pi r^2\right)^{(k-1)}}{(k-1)!} \\ &= \frac{2(\lambda\pi)^k r^{2k-1}e^{-\lambda\pi r^2}}{(k-1)!}, \end{aligned} \tag{25}$$

which generalizes (12). In other words, the probability that the $k$th closest tree to the sample point lies in the interval between $a$ and $b$ is

$$\int_a^b \frac{2(\lambda\pi)^k r^{2k-1}e^{-\lambda\pi r^2}}{(k-1)!}\,dr. \tag{26}$$

If we use the substitution $y = 2\lambda\pi r^2$ and $dy = 4\lambda\pi r\,dr$, then (26) becomes

$$\int_{2\lambda\pi a^2}^{2\lambda\pi b^2} \frac{\left(\frac{1}{2}\right)^k y^{k-1}e^{-y/2}}{(k-1)!}\,dy.$$

In other words, the pdf for $y$ is

$$g_k(y) = \frac{\left(\frac{1}{2}\right)^k y^{k-1}e^{-y/2}}{(k-1)!}$$

and so it follows from (22) that

$$2\lambda\pi R^2_{(k)} \sim \chi^2_{(2k)}. \tag{27}$$

Now assume that $n$ independent sample points are chosen at random. Similar to the derivation of (20), we have that an unbiased estimate of the density is

$$\hat{\lambda} = \frac{kn-1}{\pi\sum_{i=1}^n R^2_{(k)i}}. \tag{28}$$

Moreover, from (27) it follows that

$$2\lambda\pi\sum_{i=1}^n R^2_{(k)i} \sim \chi^2_{(2kn)}. \tag{29}$$

Consequently, a $(1-\alpha)100\%$ confidence interval for $\lambda$ is determined by the inequalities

$$\chi_{\frac{\alpha}{2}(2kn)} < 2\lambda\pi\sum_{i=1}^n R^2_{(k)i} < \chi_{1-\frac{\alpha}{2}(2kn)}.$$

Solving for $\lambda$, a $(1-\alpha)100\%$ confidence interval is given by $(C_1, C_2)$, where

$$C_1 = \frac{\chi_{\frac{\alpha}{2}(2kn)}}{2\pi\sum_{i=1}^n R^2_{(k)i}} \qquad \text{and} \qquad C_2 = \frac{\chi_{1-\frac{\alpha}{2}(2kn)}}{2\pi\sum_{i=1}^n R^2_{(k)i}}. \tag{30}$$

**A special case.** Notice that when $k = 1$ only the nearest individual to the sample point is being measured. This is the same as taking only $q = 1$ sector (the entire circle) in the two preceding sections. In particular, when $k = q = 1$, the unbiased estimates for $\lambda$ in (28) and (20) agree as do the confidence interval limits in (30) and Formula B.1.

**EXAMPLE B.2.** Use the closest trees to the 15 sample points in Example 5.2 to estimate the density and find a 95% confidence interval for this estimate.

**SOLUTION.** From Example 5.2 we have

| | | | | | | | | | | | | | | | |
|---|---|---|---|---|---|---|---|---|---|---|---|---|---|---|---|
| $R_i$ | 1.2 | 0.7 | 2.3 | 1.0 | 0.9 | 0.7 | 0.7 | 1.1 | 1.0 | 0.7 | 1.0 | 0.6 | 0.2 | 1.1 | 1.2 |
| $\pi R^2_{(1)i}$ | 4.52 | 1.54 | 16.62 | 3.14 | 2.54 | 1.54 | 1.54 | 3.80 | 3.14 | 1.54 | 3.14 | 1.13 | 0.13 | 3.80 | 4.52 |

Check that $\pi \sum_{i=1}^{15} R^2_{(1)i} = 52.64$. Since $n = 15$ and $k = 1$, then from (28)

$$\hat{\lambda} = \frac{kn - 1}{\pi \sum_{i=1}^{n} R^2_{(1)i}} = \frac{1(15) - 1}{52.64} = 0.2660 \text{ trees/m}^2$$

or 2660 trees/ha. From (30) we find

$$C_1 = \frac{\chi_{\frac{\alpha}{2}(2kn)}}{2\pi \sum_{i=1}^{n} R^2_{(1)i}} = \frac{\chi_{0.025(30)}}{2(52.64)} = \frac{16.2}{105.28} = 0.1596$$

and

$$C_2 = \frac{\chi_{1-\frac{\alpha}{2}(2kn)}}{2\pi \sum_{i=1}^{n} R^2_{(1)i}} = \frac{\chi_{0.975(30)}}{2(52.64)} = \frac{47.0}{105.28} = 0.4464.$$

This is equivalent to a confidence interval of $(1596, 4464)$ trees/ha. With fewer estimates this confidence interval is wider than the one originally calculated in Example 5.2.

### Normal Approximation

For larger values of $2kn$, one can use a normal approximation. In the case at hand, $2\lambda\pi \sum_{i=1}^{n} R^2_{(k)i} \sim \chi^2_{(2kn)}$. Adapting the argument that precedes Formula B.2 the endpoints for a $(1-\alpha)100\%$ confidence interval for $\lambda$ are determined as follows:

$$z_{\alpha/2} < \sqrt{2\left(2\lambda\pi \sum_{i=1}^{n} R^2_{(k)i}\right)} - \sqrt{2(2kn) - 1} < z_{1-\alpha/2}$$

$$\iff z_{\alpha/2} + \sqrt{4kn - 1} < \sqrt{4\lambda\pi \sum_{i=1}^{n} R^2_{(k)i}} < z_{1-\alpha/2} + \sqrt{4kn - 1}$$

$$\iff \frac{z_{\alpha/2} + \sqrt{4kn - 1}}{\sqrt{4\pi \sum_{i=1}^{n} R^2_{(k)i}}} < \sqrt{\lambda} < \frac{z_{1-\alpha/2} + \sqrt{4kn - 1}}{\sqrt{4\pi \sum_{i=1}^{n} R^2_{(k)i}}}$$

Squaring, we find that the endpoints of a $(1-\alpha)100\%$ confidence interval for $\lambda$ are

$$C_1 = \frac{\left(z_{\alpha/2} + \sqrt{4kn - 1}\right)^2}{4\pi \sum_{i=1}^{n} R^2_{(k)i}} \qquad \text{and} \qquad C_2 = \frac{\left(z_{1-\alpha/2} + \sqrt{4kn - 1}\right)^2}{4\pi \sum_{i=1}^{n} R^2_{(k)i}}. \tag{31}$$

Typically, $kn > 30$ before one would use a normal approximation.

Again note that when $k = q = 1$, (31) and Formula B.2 agree. For comparison purposes only, we now use (31) to determine a 95% confidence interval for the density in Example B.2. We obtain

$$C_1 = \frac{\left(z_{0.025} + \sqrt{4kn - 1}\right)^2}{4\pi \sum_{i=1}^{n} R^2_{(k)i}} = \frac{\left(-1.96 + \sqrt{60 - 1}\right)^2}{4(52.64)} = 0.1554$$

$$C_2 = \frac{\left(z_{0.975} + \sqrt{4kn - 1}\right)^2}{4\pi \sum_{i=1}^{n} R^2_{(k)i}} = \frac{\left(1.96 + \sqrt{60 - 1}\right)^2}{4(52.64)} = 0.4414,$$

or $(1554, 4414)$ trees/ha. This is not that different from the interval calculated in Example B.2

## Angle-Order Methods

The angle and order methods may be combined by dividing the region about each sampling point into $q$ equiangular sectors and recording the distance to the $k$th nearest individual in each sector. Morisita (1957) seems to have been the first to propose such a method.

### Randomly Distributed Populations

Though Morisita's paper concerns methods to estimate the density of non-randomly distributed populations, he first describes a combination of the angle and order methods that may be used to estimate density when a population is randomly distributed. (See Morisita (1957, equation (15) of the translation).) Assume that the distances to the $k$th closest individual in each of the $q$ sectors are measured at each sample point. In particular, let Let $R_{(k)ij}$ denote the distance from the $i$th sample point to the $k$th closest individual in the $j$th sector. Then Morisita (1957, equation 15) shows that unbiased estimate of the density[6] is

$$\hat{\lambda} = \frac{q(kqn-1)}{\pi \sum_{i=1}^{n} \sum_{j=1}^{q} R^2_{(k)ij}} \tag{32}$$

with variance

$$\operatorname{Var}(\hat{\lambda}) = \frac{\hat{\lambda}^2}{\pi^2(kqn-2)}. \tag{33}$$

Note: If $k = 1$, then this estimate for $\hat{\lambda}$ is the same as (20). Instead, if $q = 1$, then this estimate is the same as (28). And if $k = q = 1$, then this estimate is using the closest individual at each sample point and is (21)

### Non-Randomly Distributed Populations

When populations are not randomly distributed, Morisita (1957) proposed two unbiased estimates of the density and gave guidelines governing when and how each estimate was to be used. Paraphrasing Morisita, to deal with aggregated (clumped) populations, divide the total area into subplots with areas sufficiently small so that no further aggregation exists in each. Then within each subplot, either the population is distributed (1) randomly, possibly with different densities for each subplot; or (2) uniformly, again possibly with different densities for each subplot.

For the random case, Morisita showed that an unbiased estimate of the overall density is given by

$$\hat{\lambda}_1 = \frac{k-1}{\pi n} \sum_{i=1}^{n} \sum_{j=1}^{q} \frac{1}{R^2_{(k)ij}}. \tag{34}$$

Note: It is clear from the formula above that $k$ must be at least 2. Though not apparent from the formula, the estimate $\hat{\lambda}_1$ actually requires measuring the distances to the $k$th closest individual in each sector, where $k \geq 3$. This follows from the complicated formula for the variance (see Morisita (1957)) that involves a factor of $k-2$ in the denominator. This estimate is also discussed by Eberhardt (1967) and Seber (1982).

For the more uniform case, Morisita showed that an unbiased estimate of the density is given by

$$\hat{\lambda}_2 = \frac{kq-1}{\pi n} \sum_{i=1}^{n} \frac{q}{\sum_{j=1}^{q} R^2_{(k)ij}} = \frac{q(kq-1)}{\pi n} \sum_{i=1}^{n} \frac{1}{\sum_{j=1}^{q} R^2_{(k)ij}}. \tag{35}$$

Not apparent from the formula above, the estimate $\hat{\lambda}_2$ requires that $kq \geq 3$ because the variance (see Morisita (1957)) involves a factor of $kq - 2$ in the denominator.

Be careful to note the difference in order of operations (reciprocals and summations) in these two estimates. In particular,

$$\sum_{i=1}^{n} \frac{1}{\sum_{j=1}^{q} R^2_{(k)ij}} \neq \sum_{i=1}^{n} \sum_{j=1}^{q} \frac{1}{R^2_{(k)ij}} \quad \text{and} \quad \sum_{i=1}^{n} \sum_{j=1}^{q} \frac{1}{R^2_{(k)ij}} \neq \frac{1}{\sum_{i=1}^{n} \sum_{j=1}^{q} R^2_{(k)ij}}.$$

[6] In the original Japanese manuscript, equation 15 is missing a factor of $q$ and is stated incorrectly, in the notation of this paper, as

$$\hat{\lambda} = \frac{kqn-1}{\pi \sum_{i=1}^{n} \sum_{j=1}^{q} R^2_{(k)ij}}.$$

The English translation corrects this to the formula for $\hat{\lambda}$ in (32).

Notice that (35) is valid for $q = 4$ and $k = 1$, which corresponds to the using data collected in the 'standard' point-centered quarter method, and in that case simplifies to

$$\hat{\lambda}_2 = \frac{12}{\pi n} \sum_{i=1}^{n} \frac{1}{\sum_{j=1}^{4} R_{ij}^2}. \tag{36}$$

This equation differs from the earlier biased estimate of $\lambda$ for the point-centered quarter method in (1) and the unbiased estimate in Formula 5.1. Equation (36) appears to have been rediscovered by Jost (1993).

### Guidelines

Morisita (1957) summarizes the difference between the estimates as follows. In (34), the density in each sector is estimated first, by computing $\frac{1}{\pi R_{(k)ij}^2}$, and then the density of the total area is estimated using the mean of the densities of all of the sectors of all of the sample points. In contrast, in (35), the density at each sample point is calculated first, by computing $\frac{q}{\pi \sum_{j=1}^{q} R_{ij}^2}$, and then the density of the total area is estimated using the mean of densities of all of the sample points. %Thus, the former would be more appropriate if there were a significant difference %between sector densities at individual sample points, as might be the case when the distribution is random.

At first it appears that one needs to know a lot about the distribution of the population before applying one of these methods. However, Morisita makes the following observations. When the distribution is more uniform, the variance in $r$ is smaller than expected. So for two populations with the same density $\lambda$, the value of $\sum_{j=1}^{q} R_j^2$ for a uniform distribution is smaller than expected for a random distribution. (The sum of squares is minimized when all the values are the same.) The reciprocal of this sum appears in (35) for $\hat{\lambda}_2$, which tends to make $\hat{\lambda}_2$ an overestimate of the true density in the uniform case.

Similar reasoning applies to the sum of squares of the reciprocals: $\sum_{j=1}^{q} \frac{1}{R_j^2}$, too, will be smaller for a uniform distribution than expected. Since exactly this sum appears in (34) for $\hat{\lambda}_1$, it will tend to make $\hat{\lambda}_1$ an underestimate of the true density in the uniform case. Thus, for the uniform case, averaging $\hat{\lambda}_1$ and $\hat{\lambda}_2$ should provide a reasonable estimate.

Consequently, Morisita (1957) suggests the following "practical protocol" to determine which density estimate is most appropriate, even when the distribution of the population is unknown. Assuming that $k \geq 3$ (so that $\hat{\lambda}_1$ may be computed)

1. Calculate both $\hat{\lambda}_1$ and $\hat{\lambda}_2$.
2. If $\hat{\lambda}_1 \leq \hat{\lambda}_2$, use their average
$$\hat{\lambda}_0 = \frac{\hat{\lambda}_1 + \hat{\lambda}_2}{2} \tag{37}$$
as the density estimate. (The distribution may be uniform, at least locally.)
3. If $\hat{\lambda}_1 > \hat{\lambda}_2$, use $\hat{\lambda}_1$ as the density estimate. (The distribution may be random, at least locally.)

**Derivation of Equation** (35). Given our previous work, it is relatively easy to derive (35) for the case $k = 1$, that is, measuring the closest individual to the sample point in each sector (quarter). The motivating idea is to estimate the density at each point along the transect separately and then average these estimates. As usual, the density is measured by taking the reciprocal of the mean area occupied by individuals near each sample point. With $k = 1$, the mean of the $q$ estimates of the area occupied by an individual near the $i$th sample point is

$$\frac{\sum_{j=1}^{q} \pi q^{-1} R_{ij}^2}{q}.$$

The reciprocal gives an estimate of the density (near the $i$th point):

$$\frac{q}{\pi q^{-1} \sum_{j=1}^{q} R_{ij}^2}.$$

Averaging all $n$ density estimates along the transect, yields the estimate

$$\frac{1}{n}\sum_{i=1}^{n}\frac{q}{\pi q^{-1}\sum_{j=1}^{q}R_{ij}^2}.$$

However, using (19), we find that

$$\begin{aligned}
E\left[\frac{1}{n}\sum_{i=1}^{n}\frac{q}{\pi q^{-1}\sum_{j=1}^{q}R_{ij}^2}\right] &= \frac{1}{n}\sum_{i=1}^{n}E\left[\frac{q}{\pi q^{-1}\sum_{j=1}^{q}R_{ij}^2}\right]\\
&= \frac{1}{n}\sum_{i=1}^{n}\left[\int_0^\infty\cdots\int_0^\infty\frac{q}{\pi q^{-1}\sum_{j=1}^{q}R_{ij}^2}(2\lambda\pi q^{-1})^q(R_{i1}\cdots R_{iq})e^{-\lambda\pi q^{-1}\sum_{j=1}^{q}R_{ij}^2}\,dR_{i1}\cdots dR_{iq}\right]\\
&= \frac{1}{n}\sum_{i=1}^{n}\frac{\lambda q}{q-1}\\
&= \frac{\lambda q}{q-1},
\end{aligned}$$

which means that the estimate is biased. An unbiased estimate of the density is

$$\hat{\lambda} = \frac{q-1}{q}\left[\frac{1}{n}\sum_{i=1}^{n}\frac{q}{\pi q^{-1}\sum_{j=1}^{q}R_{ij}^2}\right] = \frac{q-1}{n}\sum_{i=1}^{n}\frac{q}{\pi\sum_{j=1}^{q}R_{ij}^2}.$$

This is the same as (35) with $k = 1$ or (36) with $q = 4$.

**EXAMPLE B.3.** If we use (36) and the data in Example 5.2 (where $k = 1$) we obtain

$$\hat{\lambda}_2 = \frac{12}{15\pi}\sum_{i=1}^{15}\frac{1}{\sum_{j=1}^{4}R_{ij}^2} = 0.2078\ \text{trees/m}^2. \tag{38}$$

Table 13 compares this estimate to the estimates with the other applicable methods in this paper. In short, though most estimates are similar, it is important to specify which formula one is using to estimate density when the point-centered quarter method is employed.

**TABLE 13.** The various density estimates using the data in Example 5.2.

| Equation | Formula | $\hat{\lambda}$ | Source |
|---|---|---|---|
| Equation (1) | $\frac{1}{\bar{r}^2} = \frac{16n^2}{\left(\sum_{i=1}^{n}\sum_{j=1}^{4}R_{ij}\right)^2}$ | 0.2201 | Cottam et al. (1953), Morisita (1954) |
| Formula 5.1 | $\frac{4(4n-1)}{\pi\sum_{i=1}^{n}\sum_{j=1}^{4}R_{ij}^2}$ | 0.2161 | Pollard (1971); Seber (1982) |
| Equation (28) | $\frac{kn-1}{\pi\sum_{i=1}^{n}R_{(k)i}^2}$ | 0.2660 | Pollard (1971) |
| Equation (36) | $\frac{12}{\pi n}\sum_{i=1}^{n}\frac{1}{\sum_{j=1}^{4}R_{ij}^2}$ | 0.2078 | Morisita (1957) |

Engeman et al. (1994) examined a large number of methods to estimate density[7] including those suggested above in (28), (34), and (35). Of the estimators discussed in this paper, they concluded that the best performing ones were the angle-order methods with $q = 4$ (that is, quarters) and $k = 3$ followed by $q = 4$ and $k = 2$ and then the two order methods with $k = 3$ and then $k = 2$. However, notice that the efficiency is decreased in the angle-order methods since in the first case 12 trees must be located at each sample point and in the second case 8 trees.

More recently, Khan et al. (2016) examined Morisita's angle-order estimate in equation (32) with $q = 4$ sectors and $k = 1$, 2, and 3. With these parameters, (32) simplifies to

$$\hat{\lambda}_1 = \frac{4(4n-1)}{\pi \sum_{i=1}^{n} \sum_{j=1}^{q} R_{(1)ij}^2}, \qquad \hat{\lambda}_2 = \frac{4(8n-1)}{\pi \sum_{i=1}^{n} \sum_{j=1}^{q} R_{(2)ij}^2}, \qquad \hat{\lambda}_3 = \frac{4(12n-1)}{\pi \sum_{i=1}^{n} \sum_{j=1}^{q} R_{(3)ij}^2}$$

for $k = 1$, 2, and 3, respectively. They found that these estimates "show[ed] no significant differences among them as tested in various spatial patterns except in plant assemblages with a strong repulsion (plant competition)." In this later case, the estimate with $k = 3$ performed best. Since the spatial pattern is usually unknown prior to surveying, they recommend using $k = 3$, unless plants are sparsely distributed.

# C A Non-parametric Estimate

Most of the distance method density estimates discussed so far have the disadvantage of assuming complete spatial randomness (CSR) in the distribution of the population in the area sampled. This assumption justifies the use of the Poisson distribution in developing the various density estimates. However, many authors (e.g., see Engeman et al. (1994)) suggest that plant distributions are seldom random and are often aggregated. In contrast, the use of non-parametric statistics to develop a density estimate would require no assumption about the underlying distribution of organisms.

Patil et al. (1979) and Patil et al. (1982) developed a distance-based, non-parametric estimate of plant density. It is beyond the scope of this paper to derive these formulæ. The latter paper revises their earlier work and the estimates, which we state without proof, come from the suggested formulæ in Patil et al. (1982).

### Non-parametric Estimates

Data are collected as in the special case of the order method described above. That is, at each of the $n$ sample points along the transect, the distance to the closest individual is recorded (there are no quarters). These $n$ distances are then ordered from smallest to largest. Let $R_{(k)}$ denote the $k$th order statistic, that is, the $k$th smallest such distance. Next, for any real number $r$, let $\lfloor r \rfloor$ denote the floor function, that is, the greatest integer less than or equal to $r$. Then

$$\hat{\lambda} = \frac{n^{2/3} - 1}{n\pi R^2_{(\lfloor n^{2/3} \rfloor)}}. \tag{40}$$

[7] A word of caution: The formulæ for angle-order density estimates are notationally complex and it is easy to make an error in copying or using them. The formula for the angle-order method density estimate $\hat{\lambda}_2$ in (35), which is really Morisita (1957, equation 31), is given in Engeman et al. (1994, pages 1771 and 1773) as

$$[nq(kq-1)/\pi]\, \Sigma\, 1/R^2_{(k)ij}. \tag{39}$$

That is, they multiply by $n$ rather than divide by $n$ in the first factor. Based on (39), they then write (36) as

$$\text{AO1Q:} \qquad [12n/\pi]\, \Sigma\, 1/R^2_{(1)ij},$$

and for $k = 2$ and $k = 3$, they write

$$\text{AO2Q:} \qquad [28n/\pi]\, \Sigma\, 1/R^2_{(1)ij},$$

$$\text{AO3Q:} \qquad [44n/\pi]\, \Sigma\, 1/R^2_{(1)ij}.$$

The error for AO2Q and AO3Q occurs again in White et al. (2008, Table 1).

An estimate of the variance is given by

$$\operatorname{Var}(\hat{\lambda}) = \frac{\hat{\lambda}^2}{n^{2/3}} \tag{41}$$

and so the the standard deviation is $\frac{\hat{\lambda}}{n^{1/3}}$. For large samples, a confidence interval is developed in the usual way: The endpoints of a $(1-\alpha)100\%$ confidence interval for the density $\lambda$ are well-approximated by

$$C_1 = \hat{\lambda} + \frac{z_{\frac{\alpha}{2}}\hat{\lambda}}{n^{1/3}} \qquad \text{and} \qquad C_2 = \hat{\lambda} + \frac{z_{1-\frac{\alpha}{2}}\hat{\lambda}}{n^{1/3}}. \tag{42}$$

**EXAMPLE C.1.** The data in Example B.2 lists the distances to the closest trees at $n = 15$ sample points. Estimate the density non-parametrically.
**SOLUTION.** The ordered data are

| | | | | | | | | | | | | | | | |
|---|---|---|---|---|---|---|---|---|---|---|---|---|---|---|---|
| $R_{(k)}$ | 0.2 | 0.6 | 0.7 | 0.7 | 0.7 | **0.7** | 0.9 | 1.0 | 1.0 | 1.0 | 1.1 | 1.1 | 1.2 | 1.2 | 2.3 |
| $\pi R^2_{(k)}$ | 0.13 | 1.13 | 1.54 | 1.54 | 1.54 | **1.54** | 2.54 | 3.14 | 3.14 | 3.14 | 3.80 | 3.80 | 4.52 | 4.52 | 16.62 |

Now use (40) and (41). Since $[n^{2/3}] = [15^{2/3}] = [6.08] = 6$, then $R_{([15^{2/3}])} = R_{(6)} = 0.7$. Thus,

$$\hat{\lambda} = \frac{n^{2/3} - 1}{n\pi R^2_{(\lfloor n^{2/3} \rfloor)}} = \frac{15^{2/3} - 1}{15(1.54)} = 0.2201 \text{ trees/m}^2.$$

This estimate of $\lambda$ compares favorably with those given by the parametric formulæ in Table 13.

An estimate of the variance is given by

$$\operatorname{Var}(\hat{\lambda}) = \frac{\hat{\lambda}^2}{n^{2/3}} = \frac{(0.2201)^2}{15^{2/3}} = 0.0080$$

and for the standard deviation by $\sqrt{\operatorname{Var}(\hat{\lambda})} = \sqrt{0.0080} = 0.0894$. Though the sample size is not large, we illustrate the calculation of a 95% confidence interval for $\lambda$.

$$C_1 = \hat{\lambda} + \frac{z_{0.025}\hat{\lambda}}{n^{1/3}} = 0.2201 - 1.96(0.0894) = 0.0449 \text{ trees/m}^2$$

and

$$C_2 = \hat{\lambda} + \frac{z_{0.975}\hat{\lambda}}{n^{1/3}} = 0.2201 + 1.96(0.0894) = 0.3953 \text{ trees/m}^2.$$

This confidence interval is wider than the one calculated in Example B.2 using parametric methods. In the discussion section of Patil et al. (1982), the authors note that the price for a robust density estimate "is the considerable increase in variance as compared to a parametric estimator which assumes a specific spatial distribution of plants."

## Truncated Sampling

For truncated sampling, that is, when a consistent upper limit is placed on the search radius used about each sample point, Patil et al. (1979) derived formulæ for the density and its variance. Using these formulæ with the modifications in Patil et al. (1982) leads to the following. Let $w$ be the upper limit for the radius beyond which one does not search. Let $n$ be the number of sample points and let $n_1$ denote the number of sample points with observations, that is, points where the distance to the nearest individual does not exceed $w$. (So there are $n_0 = n - n_1$ sample points without observations.) The data are the order statistics $R_{(k)}$, where $k = 1, \dots, n_1$.

Then

$$\hat{\lambda}_t = \frac{n_1}{n}\left(\frac{n_1^{2/3} - 1}{n_1 \pi R^2_{(\lfloor n_1^{2/3} \rfloor)}}\right). \tag{43}$$

An estimate of the variance is given by

$$\operatorname{Var}(\hat{\lambda}_t) = \frac{\hat{\lambda}_t^2}{n_1^{2/3}} + \hat{\lambda}_t^2\left(\frac{1}{n_1} - \frac{1}{n}\right)\left(1 + \frac{1}{n_1^{2/3}}\right). \tag{44}$$

For large samples, the endpoints of a $(1-\alpha)100\%$ confidence interval for the density $\lambda$ are well-approximated by

$$C_1 = \hat{\lambda}_t + z_{\frac{\alpha}{2}}\sqrt{\mathrm{Var}(\hat{\lambda}_t)} \qquad \text{and} \qquad C_2 = \hat{\lambda}_t + z_{1-\frac{\alpha}{2}}\sqrt{\mathrm{Var}(\hat{\lambda}_t)}. \tag{45}$$

**EXAMPLE C.2.** To illustrate these calculations return once more to the data in Example C.1. Suppose that the students who collected the data only brought a 1 meter stick with them and so did not search for trees beyond a meter from each sampling point. Then the data would consist of the $n_1 = 10$ observations that were no greater than 1.0 m. Determine a density estimate non-parametrically.

**SOLUTION.** We have $n = 15$ sampling points, $\lfloor n_1^{2/3} \rfloor = \lfloor 10^{2/3} \rfloor = 4$, and from the table in Example C.1, $R_{(4)} = 0.7$. Using (43), (44) we obtain

$$\hat{\lambda}_t = \frac{n_1}{n}\left(\frac{n_1^{2/3}-1}{n_1 \pi R^2_{(\lfloor n^{2/3} \rfloor)}}\right) = \frac{10}{15}\left(\frac{10^{2/3}-1}{10\pi R^2_{(4)}}\right) = \frac{2}{3}\left(\frac{3.6416}{10\pi(0.7)^2}\right) = 0.1577 \text{ trees/m}^2$$

and

$$\begin{aligned}
\mathrm{Var}(\hat{\lambda}_t) &= \frac{\hat{\lambda}_t^2}{n_1^{2/3}} + \hat{\lambda}_t^2\left(\frac{1}{n_1} - \frac{1}{n}\right)\left(1 + \frac{1}{n_1^{2/3}}\right) \\
&= \frac{(0.1577)^2}{10^{2/3}} + (0.1577)^2\left(\frac{1}{10} - \frac{1}{15}\right)\left(1 + \frac{1}{10^{2/3}}\right) \\
&= 0.00637.
\end{aligned}$$

So a 95% confidence interval for $\lambda$ using these data would be

$$C_1 = \hat{\lambda}_t + z_{\frac{\alpha}{2}}\sqrt{\mathrm{Var}(\hat{\lambda}_t)} = 0.1577 - 1.96\sqrt{0.00637} = 0.0013 \text{ trees/m}^2$$

and

$$C_2 = \hat{\lambda}_t + z_{1-\frac{\alpha}{2}}\sqrt{\mathrm{Var}(\hat{\lambda}_t)} = 0.1577 + 1.96\sqrt{0.00637} = 0.3141 \text{ trees/m}^2.$$

# D Shiny Apps and R Scripts

This appendix provides links to three online R Shiny Apps that may be used to calculate most of the measures discussed in the preceding pages. These require no technical knowledge of R. Additionally, for those who use R, links to several scripts are provided that will allow you to calculate these same measures and more. Examples illustrate both how to use these apps and scripts and to verify that they work as expected.

## Using the Shiny Apps

I have written three Shiny apps to calculate many of the measures discussed in the preceding sections.

- PCQM Calculator for Importance Values;
- PCQM Calculator for Density Estimates;
- PCQM Calculator for Angle-Order Density Estimates.

Each app works the same way: (1) Upload a properly formatted data file; (2) select the appropriate analysis, if necessary; and (3) read the results. Each app has a link to a Help Page describing how the data file must be formatted for the app to work correctly. These directions are summarized below for each of the apps.

### App: The PCQM Calculator for Importance Values

The PCQM Calculator for Importance Values does only one thing: It calculates importance values based on the point-centered quarter method of sampling.

**Creating the data file:** Using a spreadsheet, create a comma-separated value (`.csv`) data file of your importance value observations using the format below.

- The first row of the file should consist of column headers similar to:

  `Point,Quarter,Species,Distance,DBH`

- As indicated by the column headers, subsequent rows of the file should consist of:

  1. the sample point number along the transect;
  2. the quarter number of the observation (1–4) at the sample point;
  3. the species name of the tree (plant) nearest the sample point;
  4. the distance of this tree to the sample point in **meters** (m);
  5. the diameter at breast height (DBH) of this tree in **centimeters** (cm).

  For example, for the fifth sample point and second quarter for the data in Table 1 the row would be:

  `5,2,Acacia,2.4,5`

  where the observed tree was an *Acacia* at 2.4 m from the sample point and with a DBH of 5 cm.

  **Warning:** There should be no vacant quarters, that is, every quarter should have an observation.

- After entering all the data, the entire file should look like this one for sampling done at Pauoa Flats (Mueller-Dombois and Ellenberg (1974, Table 7.4)). The first few rows of the file are shown here.

```
Sample.Pt,Qtr.No,Species,Distance.m,DBH.cm
1,1,Psidium guajava,0.7,5.5
1,2,Acacia koa,1.6,42.5
1,3,Metrosideros collina,3.5,17
1,4,Metrosideros tremuloides,2,25
2,1,Psidium guajava,1.1,4
```

**Using the app:** Use the `Browse` button of the app to navigate to and then upload your data file. The initial setting will display the first 8 rows at the head of the file to ensure you have selected the correct one. You can change this value to display more or fewer rows; negative integer values will display rows at the tail of the file.

Once the file is uploaded, the number of sample points and the absolute density (number of trees per hectare without regard to species) are automatically displayed. Also displayed is a table with the relative density, the relative cover (also known as the relative dominance), the relative frequency, and the importance value (sum of the last three items) for each species observed. The final two columns display the relative importance and the absolute density estimate for each species.

**EXAMPLE D.1.** Figure 7 illustrates the app interface for calculating importance values using the Pauoa Flats data set. Feel free to download this data set and then try out the app.

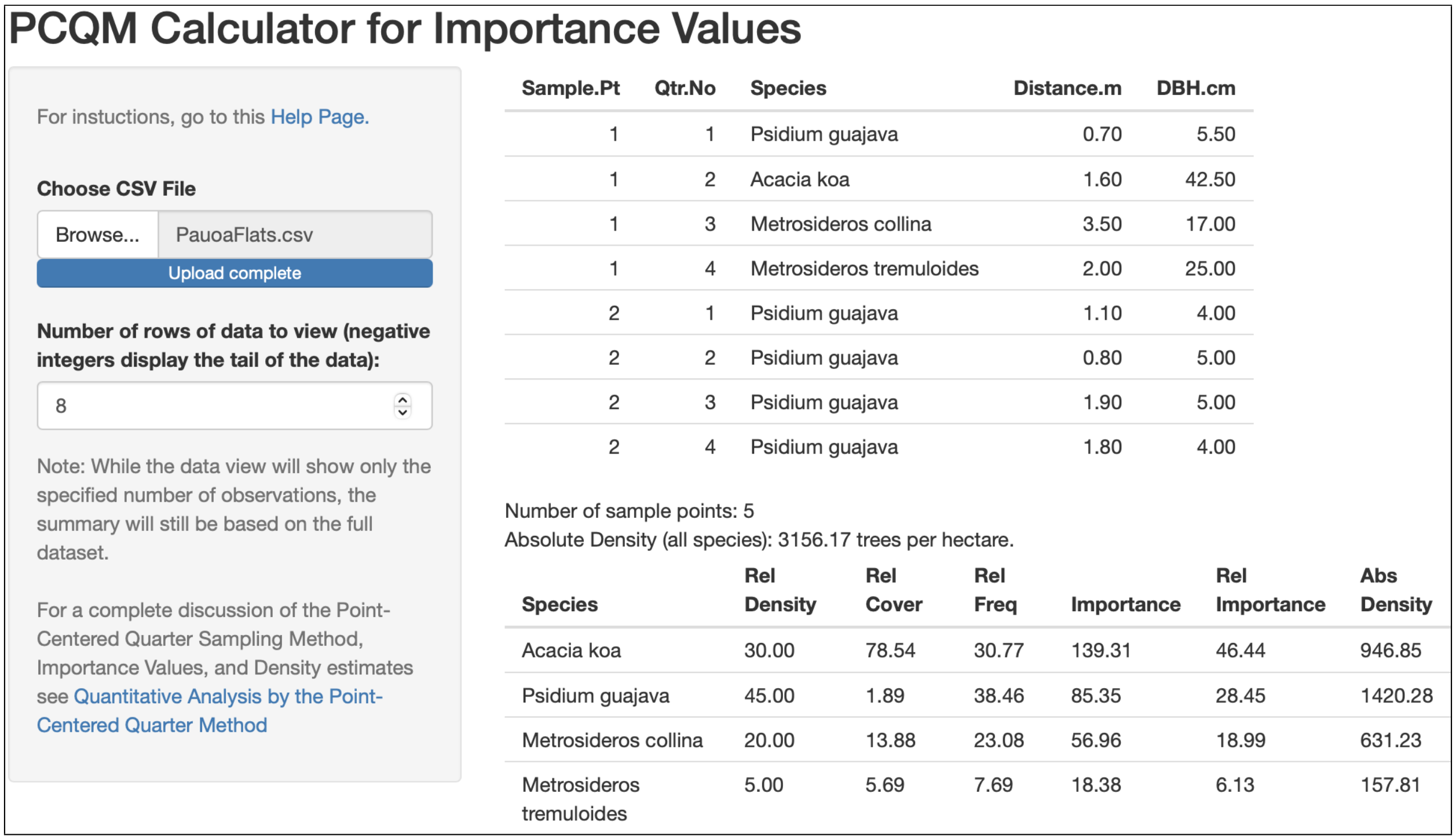

| Sample.Pt | Qtr.No | Species | Distance.m | DBH.cm |
|---|---|---|---|---|
| 1 | 1 | Psidium guajava | 0.70 | 5.50 |
| 1 | 2 | Acacia koa | 1.60 | 42.50 |
| 1 | 3 | Metrosideros collina | 3.50 | 17.00 |
| 1 | 4 | Metrosideros tremuloides | 2.00 | 25.00 |
| 2 | 1 | Psidium guajava | 1.10 | 4.00 |
| 2 | 2 | Psidium guajava | 0.80 | 5.00 |
| 2 | 3 | Psidium guajava | 1.90 | 5.00 |
| 2 | 4 | Psidium guajava | 1.80 | 4.00 |

| Species | Rel Density | Rel Cover | Rel Freq | Importance | Rel Importance | Abs Density |
|---|---|---|---|---|---|---|
| Acacia koa | 30.00 | 78.54 | 30.77 | 139.31 | 46.44 | 946.85 |
| Psidium guajava | 45.00 | 1.89 | 38.46 | 85.35 | 28.45 | 1420.28 |
| Metrosideros collina | 20.00 | 13.88 | 23.08 | 56.96 | 18.99 | 631.23 |
| Metrosideros tremuloides | 5.00 | 5.69 | 7.69 | 18.38 | 6.13 | 157.81 |

**FIGURE 7.** The PCQM Calculator for Importance Values Shiny App at https://kjmitchell.shinyapps.io/PCQM_Calculator_for_Importance_Values/.

Notice that the column headers need not be the exact names, `Point, Quarter, Species, Distance, DBH`, mentioned earlier. However, an individual header cannot not have a comma as part of its name. There is some basic error checking that occurs (e.g., every sample point must have exactly four quarters, there are no vacant cells).

The summary provides the number of sample points and the overall density. The data table lists the species in descending order of importance.

**EXERCISE D.1.** Walker (2006) sampled a *Pinus ponderosa* - *Pseudotsuga menzeisii* forest on the north-facing slope of Flagstaff Mountain, CO in 1994. (Data available at http://math.hws.edu/pcqm/FlagstaffMtn.csv.) Species key: `Pinpon` = *Pinus ponderosa*, `Psemen` = *Pseudotsuga menzeisii*.) Download the data into your own computer file and then use the app to determine the relative importance and estimated density of each species. [Answers: Relative Importance: Psemen 83.88, Pinpon 16.12. Density: Psemen 907.90, Pinpon 82.54.]

## App: PCQM Calculator for Density Estimates

The PCQM Calculator for Density Estimates may be used to calculate density estimates based on the point-centered method of sampling.

**Creating the data file:** At each sample point along the survey transect, the world is assumed to have been divided into $q$ equi-angular sectors. In each sector a distance from the sampling point to the nearest tree or plant of interest has been recorded. Most often four quarters are used at each sampling point ($q = 4$). Note:

- Four quarters ($q = 4$) are required for both the Cottam, Curtis, & Hale estimate and for the Warde & Petranka estimate. For the Pollard estimate, any number of sectors is possible.
- Additionally, the Cottam, Curtis, & Hale and the Pollard estimates require an observation in each sector. The Warde & Petranka estimate allows for vacant quarters. That is, it is designed for truncated sampling where no observation is recorded in a quarter if no tree is encountered within some fixed distance established prior to sampling.
- Finally the Patil non-parametric estimate uses a single observation (the closest tree) at each sampling point, but also allows for truncated sampling.

Create a comma-separated value data file (`.csv`) in the following format.

- The first row of the file should consist of column headers for each of the $q$ sectors. Typical headers when using quarters might be

  `Q1, Q2, Q3, Q4` or `A, B, C, D`

  or when using only $q = 2$ sectors at each sample point one might use `I, II`. **Caution:** Do not use just numeric values such as: 1, 2, 3, 4 as these values may inadvertently be treated as data.
- Subsequent rows of the file consist of the distance in **meters** (m) from the sampling point to the nearest tree in each sector. If truncated sampling is being used (Warde & Petranka or Patil estimates), some cells may be blank. Thus, a subsequent line in the data file might take the form:

  `5.4,2.1,2.2,10.8`

  where the values are the distances in meters recorded in each of the four quarters at the sample point
- After entering all the data, the entire file should look like this one for sampling done by students at Lamington National Park in Queensland.

**Using the app:** Use the `Browse` button to navigate to and then upload your data file. The initial settings will display the first 8 rows at the head of the file to ensure you have selected the correct one. You can change this value to display more or fewer rows; negative integer values will display rows at the tail of the file.

Use the radio buttons in the `Density Estimator` menu to choose the estimate appropriate to your sampling method. The number of sectors $q$ at each sampling point is automatically determined by the number of columns in the data file.

- For sampling with any number of sectors and no vacant cells, use the Pollard estimate for an unbiased density estimate with a 95% confidence interval.
- For truncated sampling (vacant cells) by quarters, use the Warde & Petranka estimate. If there are no vacant quarters, the density estimate will be the same as the Cottam, Curtis, & Hale estimate.
- For sampling by quarters with no vacant cells, the Cottam, Curtis, & Hale estimate produces a biased estimate (but has a long history of being used).
- The Patil non-parametric estimate uses just a single observation at each sampling point and also allows for truncated sampling. **Caution:** If your data file contains multiple observations at a sampling point, this method will automatically select the closest tree at each point.

**EXAMPLE D.2.** Figure 8 illustrates the app interface for calculating basic densities using the data from Example 5.2.

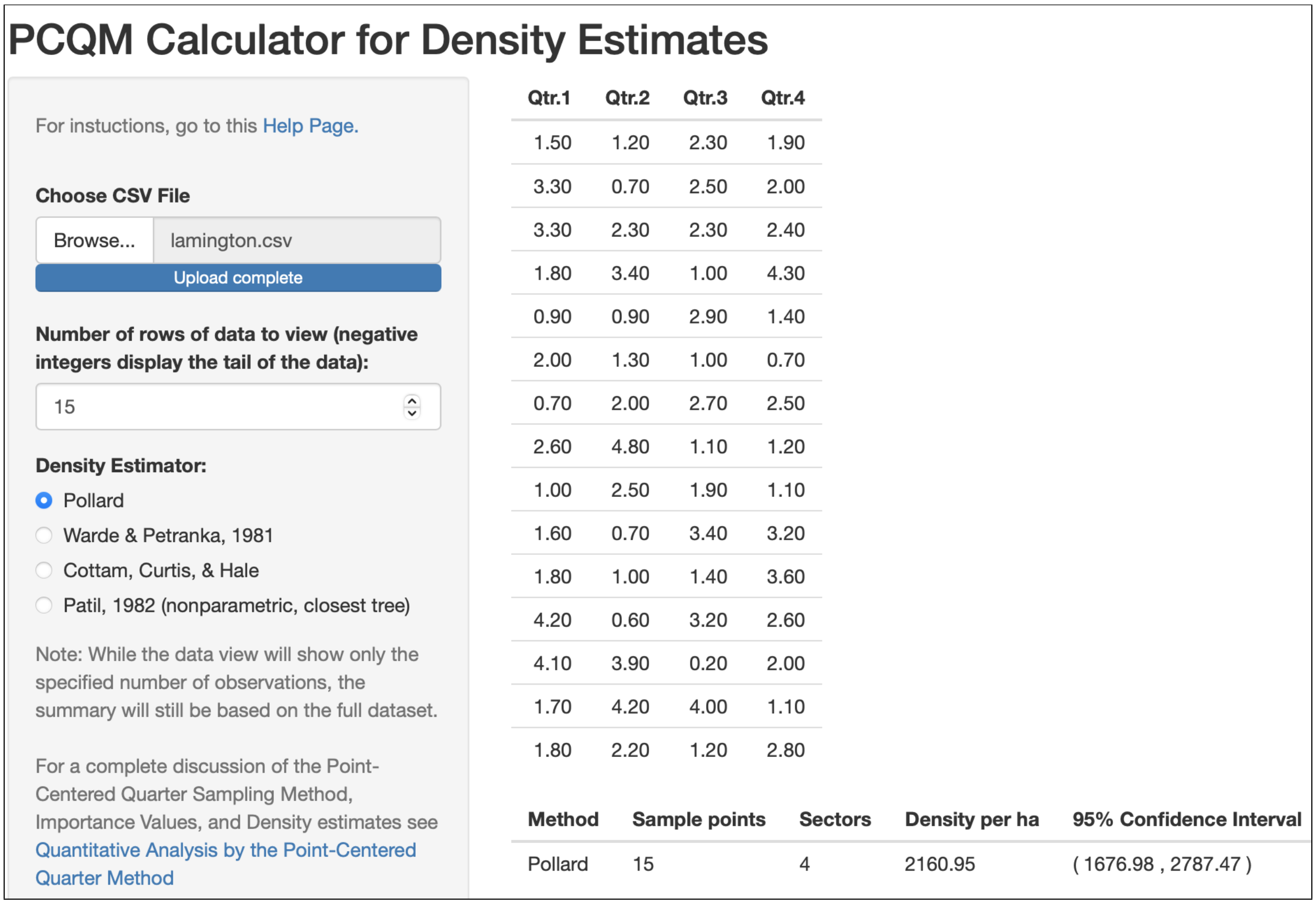

**FIGURE 8.** The PCQM Calculator for Density Estimates Shiny App at https://kjmitchell.shinyapps.io/PCQM_Calculator_for_Density_Estimates/. The summary lists the number of sample points and sectors, the density estimate, and a 95% confidence interval, if the method allows for this.

**EXERCISE D.2.** The Magellanic penguin (*Spheniscus magellanicus*) is a South American penguin, breeding in coastal Patagonia, including Argentina, Chile, and the Falkland Islands, with some migrating to Brazil and Uruguay. They are medium-sized penguins which grow to be 61–76 cm tall and weigh between 2.7 and 6.5 kg. Breeding grounds for Magellanic penguins include Isla Observatorio (shown here) and Goffré Island, Argentina in the southwestern Atlantic.

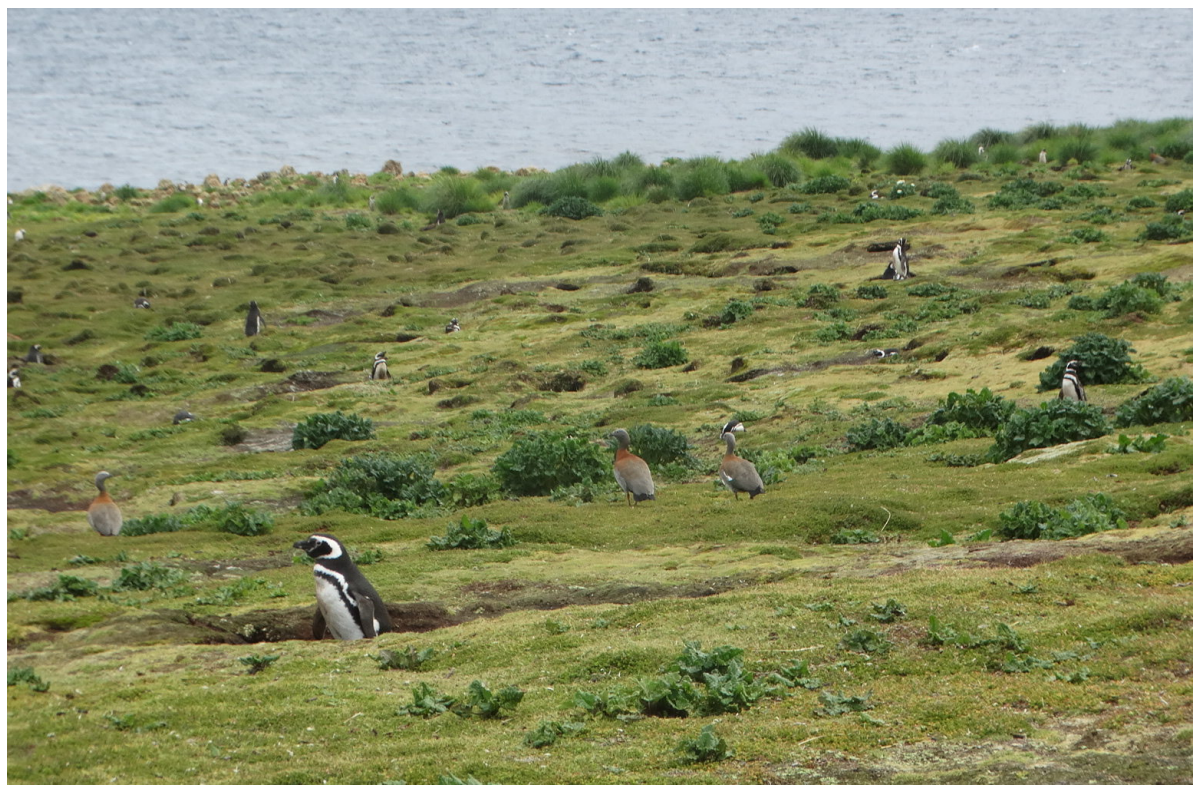

Balza et al. (2022) conducted a study to update the status of seabird colonies at both Observatorio and Goffré Islands during the breeding season. In only 4.5 km$^2$, the islands harbor roughly 90,000 breeding seabirds.

Photo of Magellanic penguins courtesy of Ulises Balza.

To estimate their numbers Balza et al. (2022) used the point-centered quarter method. Their burrowed nests require field inspection to assess occupancy. At each sample point along the transects they measured the distance to the nearest occupied nest in each of four quarters.

On Goffré Island, they observed nests only in tussock grasslands and prairie and negligible amounts in rushland. They used the Pollard method to estimate nest density.

Using their publicly available data, create a file to use with the The PCQM Calculator for Density Estimates to obtain an estimate of penguin nest density in their study area.

[Answer: 95.1 nests/hectare; confidence interval: $[72.4, 125.0]$ nests/hectare which agrees with their results.]

### App: PCQM Calculator for Density Estimates by Angle-Order Methods

The PCQM Calculator for Density Estimates by Angle-Order Methods may be used to calculate Morisita's density estimates based on angle-order methods with the point-centered method of sampling. The data file structure required to use the app is similar to that required for the app in Example D.2 above. For details see the Help Page for the app.

**EXAMPLE D.3.** Figure 9 illustrates the app interface for calculating densities using an angle-order method for the data in Example 5.2. The density estimate agrees with the value listed earlier in Table 13 for Morisita (1957) angle-order method for randomly distributed trees in (36).

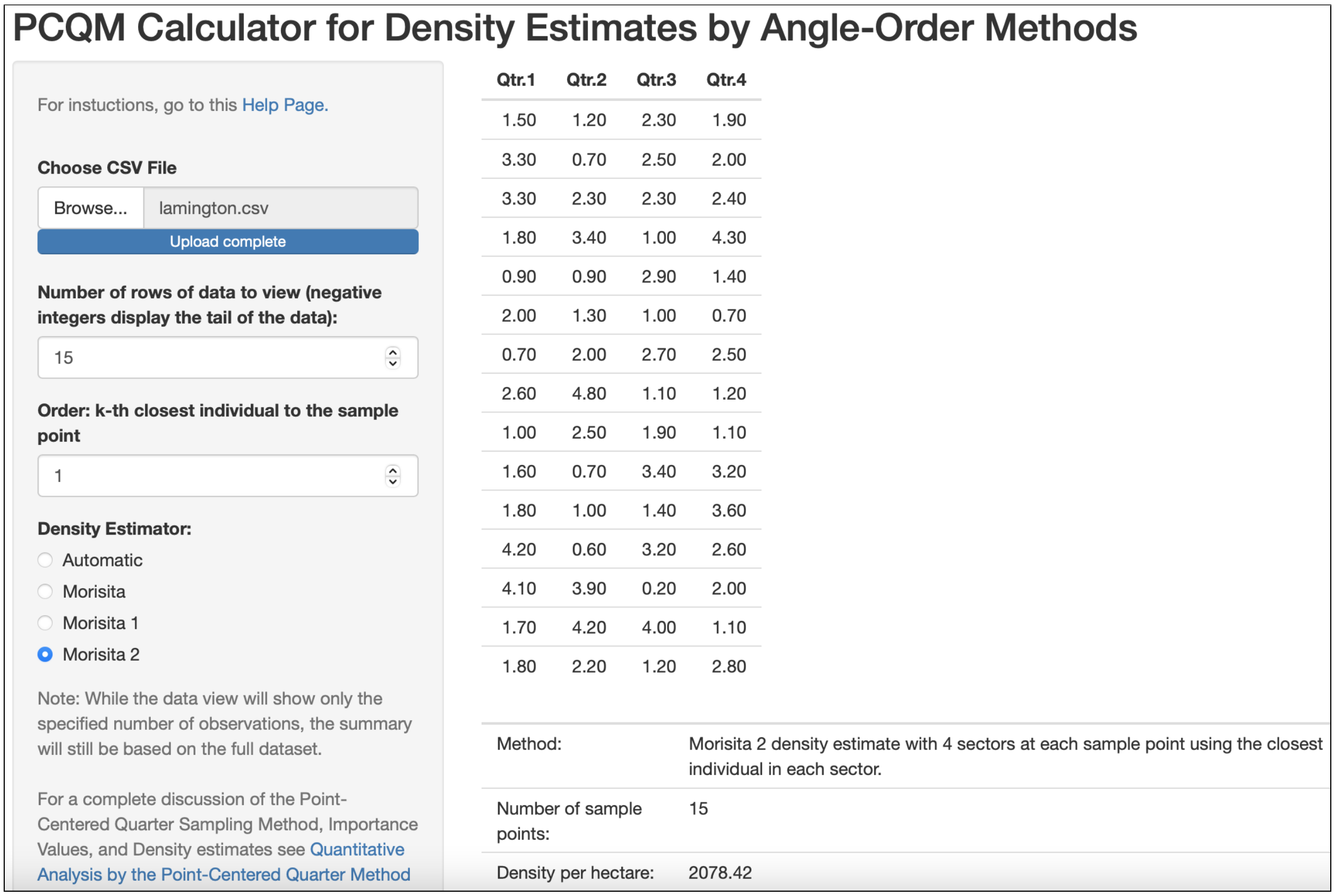

**FIGURE 9.** The PCQM Calculator for Density Estimates by Angle-Order Methods at https://kjmitchell.shinyapps.io/PCQM_Calculator_for_Angle-Order_Density_Estimates/. The summary describes the details of the method used, the number of sample points, and the overall density.

**EXERCISE D.3.** In Khan et al. (2016, Tables 1, 2, 3 in the Supplementary Materials), the authors provided three datasets from a randomly distributed simulated population with true density of 1000 trees/ha and $n = 10$ sample points. The datasets use $k = 1$, 2, and 3, that is, the first, second, and third nearest tree in each quarter.

Download their Supplementary Material: S1 Table and create data files for $k = 1$, 2, and 3 from their three tables. Use the PCQM Calculator for Density Estimates by Angle-Order Methods to carry out their analysis. Note: What Khan et al. (2016) refer to as "Corrected PCQM" is the `Morisita` option in the app. Verify that your results are identical to theirs.

## What is R?

R is a language and environment for statistical computing and graphics. It is available as free software at http://www.r-project.org under the terms of the Free Software Foundation's GNU General Public License in source code form. It compiles and runs on a wide variety of operating systems such as Windows, MacOS, and most UNIX platforms.

Initially, I used the base installation of R (and many people still do), but I now use RStudio which is an Integrated Development Environment (IDE). RStudio is a free add-on to R which provides a user-friendly interface including the R Console, a script editor, and many other features. RStudio is freely available for Windows, Mac, and Linux operating systems and can be downloaded from the RStudio site (select "Download Free Desktop IDE"). Note: You must install R before you install RStudio. You can find more information about RStudio at https://www.rstudio.com.

There are many online resources to help you learn about R. Start with the R-project homepage at http://www.r-project.org. Here are a few others written for those just beginning to use R.

1. The online text *An Introduction to* R by Douglas et al. at https://intro2r.com;
2. The RStudio Team's "RStudio Education" at https://education.rstudio.com/learn/beginner/;
3. Kelly Black's extensive "R Tutorial" at https://www.cyclismo.org/tutorial/R/index.html.

## Downloading the Basic Scripts

I have written four basic functions to carry out the calculations previously described that are contained in a single file. They may be downloaded (once per session) using R's `source( )` command as shown.

```
source("http://math.hws.edu/pcqm/pcqm.txt")
```

```
## Downloaded: importance.val( )
## Downloaded: density.est( )
## Downloaded: angle.order.est( )
## Downloaded: np.density.est( )
```

Note: Lines that begin with ## consist of output from the command that precedes them. Examples in the following sections describe what data these functions require, how this data should be formatted, how the functions are used, and verify that the functions work as expected.

## Script to Calculate Importance Value

The function `importance.val( )` may be used to calculate the importance value of various species from data collected using the point-centered quarter method. The function has the following form:

```
importance.val(z = dataframe)
```

Details: `z` is a data frame in which each row contains the following data in the order listed.

- The first column identifies the sample point along the transect. This may be a number or other descriptor.
- The second column identifies the quarter in which the observation was made. Typically quarters are designated 1, 2, 3, and 4. But other designations, such as orientation (NW, NE, SE, SW) are possible. There must be four quarters associated with each sample point and the same four quarter descriptors should (but not must) be used at each sample point.
- The third column lists the species name or other descriptor of the closest individual to the sample point in the current quarter.
- The fourth column contains the distance (in meters) of the nearest individual to the sample point in the current quarter.
- The fifth column contains a single diameter at breast height (in centimeters) of the nearest individual to the sample point in the current quarter. Multi-stem data are not allowed.

The columns may have headers. The expectation is that the data have been recorded in a spreadsheet in csv format or in a text file in five columns in the order just described. There can be no missing data (no blank cells) in the file.

The function returns the Relative Density, Relative Cover, Relative Frequency, Importance Value, and Relative Importance of each species observed along the transect(s). It also returns an estimate of the overall Absolute Density per hectare, as well as an estimate of the Absolute Density of each species. The species are listed in descending order of Importance Value.

Examples D.4 and D.5 illustrate the use of `importance.val( )` and verify its accuracy by comparing its output to calculations that appear in the literature. Note: The data files used in all of the examples are publicly accessible.

**EXAMPLE D.4.** To illustrate how calculate importance values, return to Example 7 using data from Mueller-Dombois and Ellenberg (1974), Table 7.4. These data were collected at five sample points at intervals of 5 m in a tropical insular rainforest near Honolulu, HI in 1972. The authors emphasize that these data are for illustrative purposes only and that a sample of at least 20 points per stand is recommended. The data are located in a csv file (click on the link) at http://math.hws.edu/pcqm/PauoaFlats.csv. Here are the first two lines of the data file to show the format.

```
Sample Pt,Qtr No,Species,Distance m,DBH cm
1,1,Psidium guajava,0.7,5.5
1,2,Acacia koa,1.6,42.5
```

To read and list the data with R, and then calculate the importance value of each species, use

```r
PauoaFlats <- read.csv("http://math.hws.edu/pcqm/PauoaFlats.csv", header = TRUE)
# list the data
PauoaFlats
```

| Point | Quarter | Species | Distance | DBH |
|---|---|---|---|---|
| 1 | 1 | Psidium guajava | 0.7 | 5.5 |
| 1 | 2 | Acacia koa | 1.6 | 42.5 |
| 1 | 3 | Metrosideros collina | 3.5 | 17.0 |
| 1 | 4 | Metrosideros tremuloides | 2.0 | 25.0 |
| 2 | 1 | Psidium guajava | 1.1 | 4.0 |
| 2 | 2 | Psidium guajava | 0.8 | 5.0 |
| 2 | 3 | Psidium guajava | 1.9 | 5.0 |
| 2 | 4 | Psidium guajava | 1.8 | 4.0 |
| 3 | 1 | Acacia koa | 1.3 | 75.0 |
| 3 | 2 | Psidium guajava | 0.7 | 3.0 |
| 3 | 3 | Metrosideros collina | 1.5 | 9.0 |
| 3 | 4 | Metrosideros collina | 2.0 | 23.0 |
| 4 | 1 | Acacia koa | 3.1 | 14.0 |
| 4 | 2 | Psidium guajava | 1.7 | 6.0 |
| 4 | 3 | Psidium guajava | 1.1 | 5.0 |
| 4 | 4 | Acacia koa | 1.9 | 12.0 |
| 5 | 1 | Acacia koa | 2.5 | 23.0 |
| 5 | 2 | Acacia koa | 2.2 | 18.0 |
| 5 | 3 | Psidium guajava | 1.4 | 5.0 |
| 5 | 4 | Metrosideros collina | 2.8 | 25.0 |

```r
# calculate the impartance values
importance.val(PauoaFlats)
```

```
## Number of sample points: n = 5
## Overall Absolute Density per Hectare (Cottam & Curtis): 3156.17
##
```

| | Rel Density | R Cover | R Freq | Importance | R Import | Abs Density |
|---|---|---|---|---|---|---|
| Acacia koa | 30.00 | 78.54 | 30.77 | 139.31 | 46.44 | 946.85 |
| Psidium guajava | 45.00 | 1.89 | 38.46 | 85.35 | 28.45 | 1420.28 |
| Metrosideros collina | 20.00 | 13.88 | 23.08 | 56.96 | 18.99 | 631.23 |
| Metrosideros tremuloides | 5.00 | 5.69 | 7.69 | 18.38 | 6.13 | 157.81 |

Up to rounding, the results are the same as in Mueller-Dombois and Ellenberg (1974), pages 113 and 119.

**EXAMPLE D.5.** The scripts may be used with large files. Baltzer (2007) used 105 sample points to determine importance values at Yuhas Woods, a botanically rich and diverse area in east-central Indiana.

```
Yuhas <- read.csv("http://math.hws.edu/pcqm/Baltzer2007.csv", header = TRUE)
tail(Yuhas, n = 4)      # list the observations at the last sample point
```

| | Point | Quarter | Species | Distance.m | DBH.cm |
|---|---|---|---|---|---|
| 417 | 105 | a | Acer saccharum | 0.70 | 42.6 |
| 418 | 105 | b | Prunus serotina | 7.37 | 19.0 |
| 419 | 105 | c | Prunus serotina | 9.40 | 13.8 |
| 420 | 105 | d | Quercus alba | 7.52 | 49.8 |

```
importance.val(Yuhas)
```

```
## Number of sample points: n = 105
## Overall Absolute Density per Hectare (Cottam & Curtis): 433.61
##
```

| | Rel Density | R Cover | R Freq | Importance | R Import | Abs Density |
|---|---|---|---|---|---|---|
| Acer saccharum | 31.67 | 21.17 | 23.65 | 76.49 | 25.49 | 137.31 |
| Quercus alba | 9.05 | 18.06 | 10.14 | 37.25 | 12.41 | 39.23 |
| Ulmus rubra | 9.76 | 6.90 | 11.15 | 27.81 | 9.27 | 42.33 |
| Carya ovata | 7.38 | 5.88 | 7.77 | 21.03 | 7.01 | 32.00 |
| Prunus serotina | 7.38 | 3.17 | 7.09 | 17.64 | 5.88 | 32.00 |
| Carya cordiformis | 5.95 | 5.88 | 5.41 | 17.24 | 5.75 | 25.81 |
| Aesculus glabra | 5.00 | 2.08 | 4.73 | 11.81 | 3.94 | 21.68 |
| Fraxinus pennsylvanica | 2.86 | 5.00 | 3.04 | 10.90 | 3.63 | 12.39 |
| Juglans nigra | 1.90 | 4.74 | 2.70 | 9.34 | 3.11 | 8.26 |
| Liriodendron tulipifera | 2.14 | 3.60 | 2.36 | 8.10 | 2.70 | 9.29 |
| Quercus rubra | 1.90 | 3.75 | 2.36 | 8.01 | 2.67 | 8.26 |
| Quercus muehlenbergii | 1.43 | 2.81 | 2.03 | 6.27 | 2.09 | 6.19 |
| Acer saccharinum | 1.19 | 4.10 | 0.68 | 5.97 | 1.99 | 5.16 |
| Celtis occidentalis | 2.38 | 0.88 | 2.70 | 5.96 | 1.99 | 10.32 |
| Ulmus Americana | 2.14 | 0.55 | 3.04 | 5.73 | 1.91 | 9.29 |
| Fraxinus americana | 1.90 | 1.01 | 2.70 | 5.61 | 1.87 | 8.26 |
| Carya glabra | 1.43 | 1.86 | 2.03 | 5.32 | 1.77 | 6.19 |
| Quercus palustris | 0.24 | 3.13 | 0.34 | 3.71 | 1.24 | 1.03 |
| Ostrya virginiana | 1.19 | 0.27 | 1.69 | 3.15 | 1.05 | 5.16 |
| Fagus grandifolia | 0.48 | 1.36 | 0.68 | 2.52 | 0.84 | 2.06 |
| Quercus velutina | 0.48 | 0.73 | 0.68 | 1.89 | 0.63 | 2.06 |
| Tilia americana | 0.48 | 0.35 | 0.68 | 1.51 | 0.50 | 2.06 |
| Gymnocladus dioica | 0.24 | 0.88 | 0.34 | 1.46 | 0.49 | 1.03 |
| Nyssa sylvatica | 0.24 | 0.65 | 0.34 | 1.23 | 0.41 | 1.03 |
| Acer rubra | 0.24 | 0.48 | 0.34 | 1.06 | 0.35 | 1.03 |
| Populus Deltoides | 0.24 | 0.40 | 0.34 | 0.98 | 0.33 | 1.03 |

| | Rel Density | R Cover | R Freq | Importance | R Import | Abs Density |
|---|---|---|---|---|---|---|
| Carya laciniosa | 0.24 | 0.19 | 0.34 | 0.77 | 0.26 | 1.03 |
| Cercis canadensis | 0.24 | 0.09 | 0.34 | 0.67 | 0.22 | 1.03 |
| Carpinus caroliniana | 0.24 | 0.04 | 0.34 | 0.62 | 0.21 | 1.03 |

The results agree with those reported in Baltzer (2007) and Baltzer et al. (2007).

**EXERCISE D.4.** Create a csv file for the data in Table 1. Verify that the analysis in Table 9 is correct. You will have to make one modification to Table 1. The multi-stem DBH at one of the quarters must converted to an equivalent single stem DBH of 10 cm.

**EXERCISE D.5.** Ten sample points were used to survey a forest plot in the northwestern United States. Determine the relative importance and density estimate for each of the four species in the data set at http://math.hws.edu/pcqm/fir.csv. Species key: AGBR = *Abies grandis*, PSME = *Pseudotsuga menziesii*, PIPO = *Pinus ponderosa*, and LAOC = *Larix occidentalis*. [Answer. Relative importance (in the order above): 77.17, 12.98, 6.51, and 3.34; Density: 17.80, 2.78, 1.11, 0.56.]

**EXERCISE D.6.** Analyze a large data set of 255 sample points recorded in Alaska in 2000 by Hollingsworth (2005). The data are the first 51 transects listed in http://www.lter.uaf.edu/ascii/files/138_pcqbs_ht_diam.txt. Species key: `picmar` = *Picea mariana*, `picgla` = *Picea glauca*, `alncri` = *Alnus crispa*, `betneo` = *Betula neoalaskana*, `larlar` = *Larix laricina*, and `salgla` = *Salix glauca*. Determine the relative importance and density estimate for each species. [Partial answer: Relative Importance: picmar 92.08, betneo 1.27. Density: picmar 3125.34, betneo 29.00.]

## Script to Estimate Density

The function `density.est( )` may be used to calculate densities of organisms from data collected using the point-centered quarter method. The function has the following form:

**`density.est(z = dataframe, method = c("pollard", "cottam", "warde"), conf.level = 0.95)`**

Details: `z` is a data frame where each row contains the distances from the sample point to the nearest individual in each sector. The argument `method` specifies how the the calculation is made:

`"pollard"` uses Formula 5.1 on page 12;

`"cottam"` uses the formula in equation (1) on page 7;

`"warde"` uses the formula in equation (8) on page 16.

When `method = "pollard"`, a fixed number $q$ of equiangular sectors must be used at each sample point. For methods `"cottam"` and `"warde"` quarters, that is, four equiangular sectors, are required at each sample point.

The default method is `"pollard"`. The method `"warde"` should be used for data sets containing vacant quarters, that is, quarters in which no individual was found. If methods `"pollard"` or `"cottam"` are used with missing observations (`NA`), the calculation method automatically defaults to `"warde"`, assuming that quarters were used.

`conf.level` specifies the confidence level of the interval and has a default value of `0.95`. A confidence interval is computed only when `method = "pollard"`.

**EXAMPLE D.6.** The data in Example 5.2 (see http://math.hws.edu/pcqm/lamington.csv for format) measured the distance to the nearest tree in each quarter for 15 sample points. After reading and listing the data, to estimate the density and calculate a 95% confidence interval using Pollard's method, use the `density.est( )` function with its default arguments `method = "pollard"` and `conf.level = 0.95`, which need not appear in the function call.

```
lamington <- read.csv("http://math.hws.edu/pcqm/lamington.csv")
lamington
```

| Qtr.1 | Qtr.2 | Qtr.3 | Qtr.4 |
|---|---|---|---|
| 1.5 | 1.2 | 2.3 | 1.9 |
| 3.3 | 0.7 | 2.5 | 2.0 |
| 3.3 | 2.3 | 2.3 | 2.4 |
| 1.8 | 3.4 | 1.0 | 4.3 |
| 0.9 | 0.9 | 2.9 | 1.4 |
| 2.0 | 1.3 | 1.0 | 0.7 |
| 0.7 | 2.0 | 2.7 | 2.5 |
| 2.6 | 4.8 | 1.1 | 1.2 |
| 1.0 | 2.5 | 1.9 | 1.1 |
| 1.6 | 0.7 | 3.4 | 3.2 |
| 1.8 | 1.0 | 1.4 | 3.6 |
| 4.2 | 0.6 | 3.2 | 2.6 |
| 4.1 | 3.9 | 0.2 | 2.0 |
| 1.7 | 4.2 | 4.0 | 1.1 |
| 1.8 | 2.2 | 1.2 | 2.8 |

```
density.est(lamington)
```

```
##
##  Pollard's estimate of density using 4 sectors per point
##
## data:  lamington
## No. of sample pts: n = 15
## 95 percent confidence interval:
##  1676.98 2787.47
## sample estimates:
## density per hectare
##             2160.95
```

The density estimate is identical to that calculated in Example 5.2, but the confidence interval is narrower. The confidence interval calculated by `density.est( )` is exact, using the chi-squared distribution as in Example B.1, rather than using a normal approximation. To just calculate a 99% confidence interval for the density, use `conf.level = 0.99` and display the result as a vector. The interval widens with an increase in confidence.

```
as.vector(density.est(lamington, conf.level = 0.99)$conf.int)
```

```
## [1] 1535.59 2996.91
```

To use the density estimate proposed in Cottam et al. (1953), set `method = "cottam"`.

```
density.est(lamington, method = "cottam")
```

```
##
##  Cottam, Curtis, & Hale Point-Centered Quarter Method density estimate
##
## data:  lamington
## No. of sample pts: n = 15
## sample estimates:
## density per hectare
##              2200.7
```

Suppose that we remove two of the observations from the `lamington` data set to simulate that there were two quarters in which an individual could not be found. It is now appropriate to use the method described by Warde and Petranka (1981) to estimate the density.

The data are located at http://math.hws.edu/pcqm/lamington.vacant.csv. Notice that the missing values are in rows 9 (trailing comma) and 13 (consecutive commas).

```
Qtr 1,Qtr 2,Qtr 3,Qtr 4
1.5,1.2,2.3,1.9
3.3,0.7,2.5,2
3.3,2.3,2.3,2.4
1.8,3.4,1,4.3
0.9,0.9,2.9,1.4
2,1.3,1,0.7
0.7,2,2.7,2.5
2.6,4.8,1.1,1.2
1,2.5,1.9,
1.6,0.7,3.4,3.2
1.8,1,1.4,3.6
4.2,0.6,3.2,2.6
4.1,,0.2,2
1.7,4.2,4,1.1
1.8,2.2,1.2,2.8
```

```r
lamington.vacant <- read.csv("http://math.hws.edu/pcqm/lamington.vacant.csv", header = TRUE)
# list the data; note that empty numerical cells are replaced with NA
lamington.vacant
```

| Qtr.1 | Qtr.2 | Qtr.3 | Qtr.4 |
|---|---|---|---|
| 1.5 | 1.2 | 2.3 | 1.9 |
| 3.3 | 0.7 | 2.5 | 2.0 |
| 3.3 | 2.3 | 2.3 | 2.4 |
| 1.8 | 3.4 | 1.0 | 4.3 |
| 0.9 | 0.9 | 2.9 | 1.4 |
| 2.0 | 1.3 | 1.0 | 0.7 |
| 0.7 | 2.0 | 2.7 | 2.5 |
| 2.6 | 4.8 | 1.1 | 1.2 |
| 1.0 | 2.5 | 1.9 | NA |
| 1.6 | 0.7 | 3.4 | 3.2 |
| 1.8 | 1.0 | 1.4 | 3.6 |
| 4.2 | 0.6 | 3.2 | 2.6 |
| 4.1 | NA | 0.2 | 2.0 |
| 1.7 | 4.2 | 4.0 | 1.1 |
| 1.8 | 2.2 | 1.2 | 2.8 |

```r
density.est(lamington.vacant, method = "warde")
```

```
##
##  Warde & Petranka (1981) estimate of density using 4 sample quarters per
##  point
##
## data:  lamington.vacant
## No. of sample pts: n = 15, No. of vacant quarters: = 2
## sample estimates:
## density per hectare
##             2024.03
```

**EXAMPLE D.7.** Pollard's method requires the use of equiangular sectors at each point, not necessarily quarters. In Example 5.2, that is, Example D.6, quarters 1 and 2 were on the eastern side of the transect while quarters 3 and 4 were on the western side. Instead of using quarters at each sample point, assume we divided the world into east and west halves and selected the nearest individual in each half. We can use the original data to obtain this information and then calculate the corresponding density estimate.

```
East <- apply(lamington[,1:2],1,min)
West <- apply(lamington[,3:4],1,min)
lamington.2 <- data.frame(East, West)
tail(lamington.2, n = 2)        # check that the data is as expected
```

| | East | West |
|---|---|---|
| 14 | 1.7 | 1.1 |
| 15 | 1.8 | 1.2 |

```
density.est(lamington.2)
```

```
##
##  Pollard's estimate of density using 2 sectors per point
##
## data:  lamington.2
## No. of sample pts: n = 15
## 95 percent confidence interval:
##  1414.93 2911.44
## sample estimates:
## density per hectare
##             2027.23
```

The density estimate is within 6% of the original estimate of 2160.95 trees/ha in Example D.6, but required about half the sampling effort. However, with less data, the confidence interval is wider.

**EXERCISE D.7.** A problem in Krebs (1999) asks to estimate the population density and find a 95% confidence interval for the density for Amabilis fir in the Coast Range of British Columbia using point-quarter data collected at $n = 10$ points. The data may be found at http://math.hws.edu/pcqm/Krebs.csv. Use an appropriate R function to determine the answer. [Answer: Density per hectare: 192.958; confidence interval: $[141.387, 263.780]$.]

**EXERCISE D.8.** (From Pierce et al. (2012), pp. 299–300.) To estimate the number of active bird nests in a study area during breeding season, a transect with five sample points spaced 100 m apart was used and the distances (in m) to the nearest nest in each quarter were recorded. I: 0, 15, 15, 15; II: 10, 7, 7, 5; III: 1, 12, 3, 11; IV: 10, 10, 12, 1; V: 11, 9, 9, 7. Estimate the number of nests per hectare and determine a 95% confidence interval for the density. [Answer: Density per hectare 130.77; confidence interval: $[84.08, 204.21]$.]

## Script to Estimate Density Using Angle-Order Methods

Morisita (1957) combined the angle and order methods by dividing the region about each sample point into $q$ equiangular sectors and recording the distance to the $k$th nearest individual in each sector. The function `angle.order.est( )` calculates densities of organisms from data collected in this way.

```
angle.order.est(z = dataframe, k = 3, method = c("auto", "morisita1", "morisita2",
                   "morisita"), conf.level = 0.95)
```

Details: `z` is a data frame where each row contains the $q$ observations for a single sample point. The argument `k` specifies that the observations are distances to the $k$th nearest individual in each of the $q$ equiangular sectors. No missing observations are allowed. The default is `k = 3`. The argument `method` specifies how the the calculation is made:

`"auto"` is the default method and uses Morisita's guidelines (see page 28) to estimate the density. If $\hat{\lambda}_1 > \hat{\lambda}_2$, then $\hat{\lambda}_1$ is used. Otherwise $\hat{\lambda}_0 = \dfrac{\hat{\lambda}_1 + \hat{\lambda}_2}{2}$ is used. This method requires that $k \geq 3$.

"morisita1" uses equation (34): $\hat{\lambda}_1 = \frac{k-1}{\pi n}\sum_{i=1}^{n}\sum_{j=1}^{q}\frac{1}{R^2_{(k)ij}}$. This method requires that $k \geq 3$.

"morisita2" uses equation (35): $\hat{\lambda}_2 = \frac{kq-1}{\pi n}\sum_{i=1}^{n}\frac{q}{\sum_{j=1}^{q} R^2_{(k)ij}}$. This method requires that $kq \geq 2$.

"morisita" uses equation (32): $\hat{\lambda} = \frac{q(kqn-1)}{\pi \sum_{i=1}^{n}\sum_{j=1}^{q} R^2_{(k)ij}}$. Note: If $k = 1$, then this estimate is the same as (20), which is computed when `method = "pollard"` in the function `density.est( )`. Instead, if $q = 1$, then this estimate is the same as (28).

`conf.level` specifies the confidence level of the interval and has a default value of `0.95`. A confidence interval is computed only when `method = "morisita"`.

**EXAMPLE D.8.** Redo Example B.2 that used the closest trees to the 15 sample points in Example 5.2 to estimate the density and find a 95% confidence interval for this estimate. This requires using `method = "morisita"` and `k = 1` in the `angle.order.est( )` function. Re-use the `lamington` data from Example D.6. The results, up to rounding, are the same as calculated earlier in Example B.2.

```
lamington.closest <- data.frame(min.dist = apply(lamington, 1, min))# closest individual
angle.order.est(lamington.closest, k = 1, method = "morisita")
```

```
##
##  Morisita angle-order density estimate for randomly distributed
##  populations with 1 sector at each sample point using the closest
##  individuals.
##
## data:  lamington.closest
## Number of sample points: n = 15
## 95 percent confidence interval:
##  1594.47 4461.20
## sample estimates:
## density per hectare
##             2658.91
```

**EXAMPLE D.9.** In Example B.3 we used (36) and the `lamington` data in Example 5.2 with $k = 1$ to obtain a density estimate of 2078 trees/ha. Equation (36) corresponds to `method = "morisita2"` and `k = 1`.

```
angle.order.est(lamington, k = 1, method = "morisita2")
```

```
##
##  Morisita 2 density estimate with 4 sectors at each sample point using
##  the closest individual in each sector.
##
## data:  lamington
## Number of sample points: n = 15
## sample estimates:
## density per hectare
##             2078.42
```

## Script to Estimate Density Non-parametrically

Appendix C described a distance-based, non-parametric estimate of plant density developed by Patil et al. (1982). The data for this method consist of the distances of the nearest individual to each sample point. The function `np.density.est( )` may be used to calculate densities in this way.

```
np.density.est(z = dataframe, conf.level = 0.95)
```

Details: `z` is a data frame comprising a single column of distances of the nearest individual to each sample point. Missing observations should be indicated by `NA`, though the function should automatically detect blank entries. `conf.level` specifies the confidence level of the interval and has a default value of `0.95`.

**EXAMPLE D.10.** Example C.1 used the method of Patil et al. (1982) to non-parametrically estimate density and locate a 95% confidence interval. It employed the `lamington.closest` data set from Example D.8 that listed the distances to the nearest trees from 15 sample points. The calculations below agree with those in Example C.1.

```
np.density.est(lamington.closest)
```

```
##
##  Patil, et. al. (1982) non-parametric density estimate using the closest
##  individual at each sample point.
##
## data:  lamington.closest
## No. of sample pts: n = 15
## 95 percent confidence interval:
##   451.802 3950.144
## sample estimates:
## density per hectare
##             2200.97
```

Example C.2 also used the data in Example C.1 (that is, `lamington.closest`), but omitted all five distances greater than 1.0 m. The results using the truncated data agree with those in Example C.2.

```
lamington.trunc <- lamington.closest
# substitute NA for distances greater than 1 m
lamington.trunc[lamington.trunc > 1] <- NA
np.density.est(lamington.trunc)
```

```
##
##  Patil, et. al. (1982) non-parametric density estimate using the closest
##  individual at each sample point.
##
## data:  lamington.trunc
## No. of sample pts: n = 15, No. of truncated points: = 5
## 95 percent confidence interval:
##    13.2614 3140.8984
## sample estimates:
## density per hectare
##             1577.08
```

**EXAMPLE D.11.** To illustrate their method, Patil et al. (1982) used data from a 100-point transect in Batcheler (1971). These data may be found at http://math.hws.edu/pcqm/Batcheler1971.csv. The original data were recorded in feet and must be converted to meters before calculation. Further, the density was given in acres, so the output of `np.density.est( )`, which is in hectares, must be converted to acres.

```
Batcheler <- read.csv("http://math.hws.edu/pcqm/Batcheler1971.csv", header = TRUE)
# list the last four data values
tail(Batcheler, n = 4)
```

| | Batcheler.1971 |
|---|---|
| 97 | 18.5 |
| 98 | 25.7 |
| 99 | 20.6 |
| 100 | 25.6 |

```
BatchelerMeters <- 0.30480 * Batcheler  # convert feet to meters
np.density.est(BatchelerMeters)
```

```
##
##  Patil, et. al. (1982) non-parametric density estimate using the closest
##  individual at each sample point.
##
## data:  BatchelerMeters
## No. of sample pts: n = 100
## 95 percent confidence interval:
##  200.826 494.387
## sample estimates:
## density per hectare
##             347.606
```

```
# convert to acres
cat("Density estimate:", 0.404686*np.density.est(BatchelerMeters)$est)
```

```
## Density estimate: 140.671
```

```
cat("Confidence interval:", 0.404686*np.density.est(BatchelerMeters)$conf.int)
```

```
## Confidence interval: 81.2713 200.071
```

Up to rounding, these results agree with Patil et al. (1982, page 247), where the density is given as 140 plants per acre. The standard deviation is given as 30 plants. From (42), the confidence interval limits using 140 and 30 are shown below, and agree with those calculated above, given rounding.

```
# confidence limits
140 + 1.96 * c(-30, 30)
```

```
## [1]  81.2 198.8
```

For comparison, we estimate the density with Pollard's method.

```
cat("Density estimate:", 0.404686*density.est(BatchelerMeters, method = "pollard")$est)
```

```
## Density estimate: 142.485
```

```
cat("Confidence interval:", 0.404686*density.est(BatchelerMeters, method = "pollard")$conf.int)
```

```
## Confidence interval: 117.103 173.47
```

The two density estimates differ by 1.5%. However, the 95% confidence interval is nearly twice as wide using the non-parametric method, as Patil et al. (1982) warned.

Suppose that the sampling effort is truncated with a maximum search radius of 7.8 ft. That is, if no plant is found within that radius of the sample point, then the search is terminated. (The value of 7.8 ft was used in Patil et al. (1979) because it eliminates half the data.) To carry out the analysis, first truncate the data.

```
# substitute NA if > 7.8 ft
BatchelerMeters[BatchelerMeters > 0.30480 * 7.8] <- NA
np.density.est(BatchelerMeters)
```

```
##
##  Patil, et. al. (1982) non-parametric density estimate using the closest
##  individual at each sample point.
##
## data:  BatchelerMeters
## No. of sample pts: n = 100, No. of truncated points: = 50
## 95 percent confidence interval:
##  170.299 620.798
```

```
## sample estimates:
## density per hectare
##             395.548
```

```
# convert to acres
cat("Density estimate:", 0.404686*np.density.est(BatchelerMeters)$est)
```

```
## Density estimate: 160.073
```

```
cat("Confidence interval:", 0.404686*np.density.est(BatchelerMeters)$conf.int)
```

```
## Confidence interval: 68.9176 251.228
```

The result is within 15% of the original density estimate and would have required just half the sampling effort. However, the confidence interval is about 1.5 times wider than before.

**EXERCISE D.9.** We saw in Exercise D.2 that Balza et al. (2022) used Pollard's method to estimate the density of active Magellanic penguin nests on Goffré Island on the coast of southern Argentina. There they estimated that there were

95.1 nests/hectare;

95% confidence interval: $[72.4, 125.0]$ nests/hectare.

In a companion study on neighboring Observatorio Island, for technical reasons they used Patil's method to estimate the density of active nests that island. They found that estimated Magellanic penguin densityon Observatorio Island was similar to that on Goffré Island:

80.4 nests/ha;

95% confidence interval: $[42.4, 118.5]$ nests/hectare.

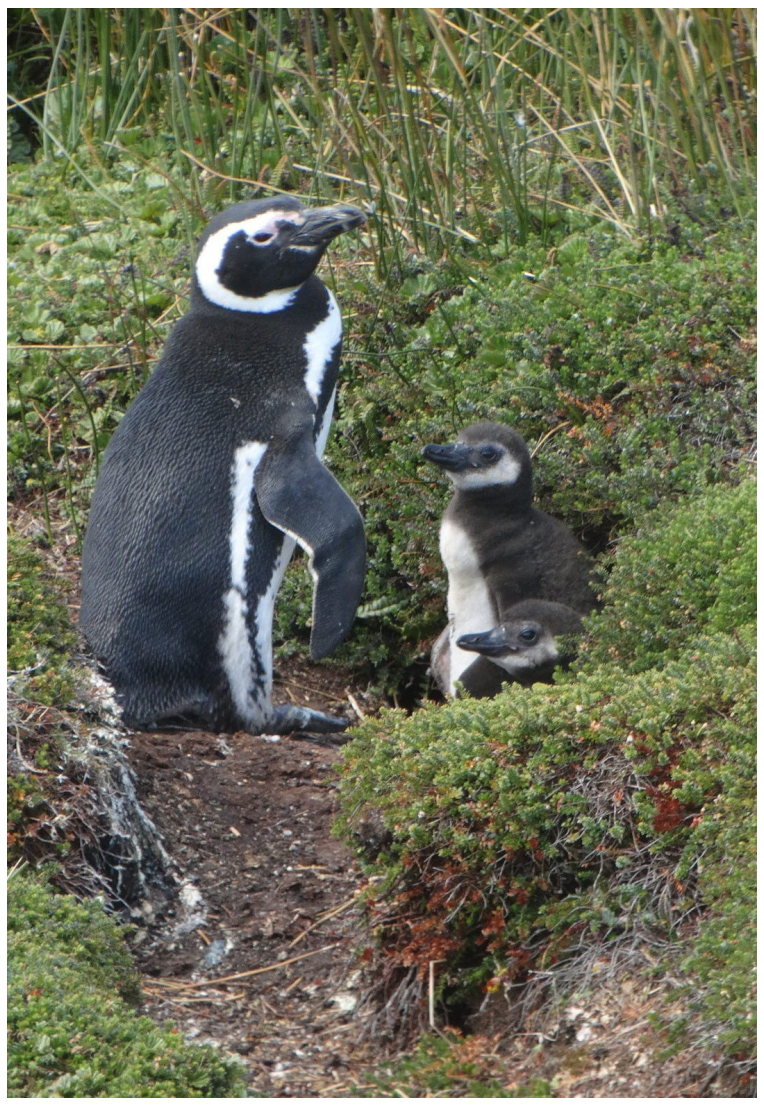

Photo of Magellanic penguins on Observatorio Island courtesy of Ulises Balza.

Balza et al. (2022) have made their data publicly available on GitHub. Carry out your own analysis of their data as follows.

- Use the function `read.csv( )` to download their data from https://raw.githubusercontent.com/ulisesbalza/magellanic_penguin_density/main/data_observatorio_np.csv and name the new file `Observatorio`
- Use the function `np.density.est( )` to determine the density.

How many points were in their data set? Do your results agree with theirs?

## Re-analyzing Data

The following functions may be used to further analyze survey data that consist of the rectangular coordinates of all the trees in a particular plot. This might be done to determine which methods provide accurate density estimates for particular forest types. These functions may be downloaded using `source("http://math.hws.edu/pcqm/survey.txt")`.

**`convex.hull(z = dataframe, lcolor = "orange", display = TRUE)`**

Details: `z` specifies a data frame with two columns containing the $x$ and $y$ coordinates, in meters, of all individuals in the area surveyed; `lcolor` specifies the color used to draw the convex hull (smallest convex polygon, i.e., "rubber band") containing the surveyed individuals, with default set to `"orange"`.

Output: The area of the convex hull and the density calculated using the hull area are reported. The area of the smallest rectangle, with sides parallel to the axes, that contains the data and the corresponding density are also reported. If `display = TRUE`, a plot of the data, its convex hull, and bounding rectangle are drawn.

**`manual.survey <- function(z = dataframe, t.coords = c(0, 0, 10, 10), points = 10,`**

```
                transects = 2, k = 1, horizontal = TRUE, display = TRUE)
```

Details: `manual.survey( )` takes the locations of all individuals in a specified area and produces the data required to estimate density using the functions described earlier in this section. `z` specifies a data frame with two columns containing the $x$ and $y$ coordinates, in meters, of all individuals in the area surveyed; `t.coords` is a vector of the form $(x_1, y_1, x_2, y_2, \dots)$ containing the coordinates of either two or four points that are used to determine either one or more transects, respectively. If four points are used, they should be listed counter-clockwise starting from the lower left. `points` specifies the number of sample points; `transects` specifies the number of transects; `k` specifies that the $k$th closest individual is sampled in each quarter with default `k = 1`; `display` is the logical value specifying whether the forest is shown, with default `TRUE`; and `horizontal` is a logical value specifying whether the orientation of the transects is horizontal (by default `TRUE`) or not (vertical).

Transects: If `t.coords` contains four components specifying two points, then these two points are used as the endpoints of a single transect with the specified number of sample points distributed evenly along its length. If `t.coords` contains eight components specifying four points and `horizontal = TRUE`, then the first two points are used as the endpoints of the bottommost transect and the remaining two points are used as the endpoints of the uppermost transect. Any remaining transects are interpolated between these two with the sample points distributed evenly along them. If `horizontal = FALSE`, the first and fourth points and second and third points determine the outermost transects. Note: The lines defining the quarters are always parallel to the axes regardless of whether the transects are.

Output: The area of the convex hull and the density calculated using the hull area are reported, as are the area of the bounding rectangle the corresponding density. A data frame comprised of the distances to the $k$th closest individual in each quarter is returned. If `display = TRUE`, a plot of the data, its convex hull and bounding rectangle, and the transects with sample points as specified are drawn, with the $k$th closest individual in each quarter highlighted.

**EXAMPLE D.12.** In a study measuring the impact of deer browsing on tree regeneration, Newell (2015) collected coordinate data for all 257 trees in a plot between Snell and Pre-Emption Roads in Geneva, NY (see http://math.hws.edu/pcqm/NewellSouth.csv). To determine the site's convex hull and bounding rectangle areas and the corresponding densities, use the `convex.hull( )` function. A plot is shown later in this example.

```
source("http://math.hws.edu/pcqm/survey.txt")
```

```
## Downloaded convex.hull( )
## Downloaded manual.survey( )
```

```
NewellSouth <- read.csv("http://math.hws.edu/pcqm/NewellSouth.csv")
convex.hull(NewellSouth, display = FALSE)
```

```
## Data: NewellSouth
##                      Area (ha) Density (per ha)
## Convex Hull              0.241          1056.00
## Bounding Rectangle       0.274           930.50
```

The actual density is known, so there would seem to be no reason to estimate it. However, if one now wanted to sample, rather than completely survey, other similar areas, then it would be useful to know which sampling techniques provide accurate estimates. One can test these techniques using the previously collected data. To place transects on the site and then see how various density estimates compare to the actual density, use the `manual.suvey( )` function. Its output can be analyzed with an appropriate density function. To illustrate the idea, we use three horizontal transects with a total of 15 sample points to which are measured the closest individuals in each quarter. We then use Pollard's method to analyze the data.

```
Newell.horiz <- manual.survey(z = NewellSouth, t.coords = c(15, 8, 65, 8, 65, 26, 15, 26),
    points = 15, transects = 3, k = 1)
```

```
## Data: NewellSouth
##                      Area (ha) Density (per ha)
```

```
## Convex Hull             0.241          1056.00
## Bounding Rectangle      0.274           930.50
```

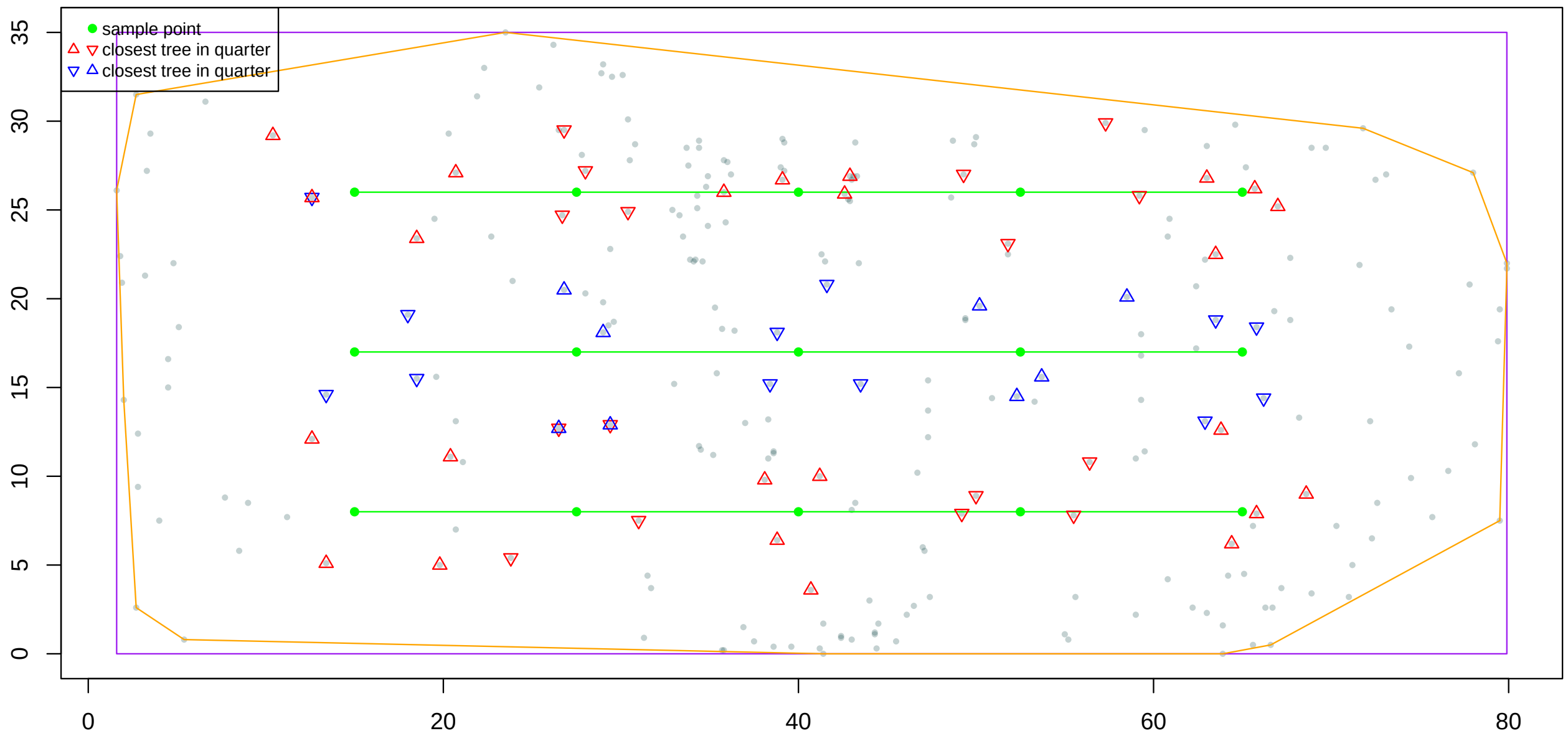

```
# data for the last three sample points
tail(Newell.horiz, n = 3)
```

| | Qtr 1 | Qtr 2 | Qtr 3 | Qtr 4 |
|---|---|---|---|---|
| 13 | 3.036445 | 1.14017 | 4.20000 | 2.60192 |
| 14 | 6.184658 | 3.35261 | 2.98329 | 6.70298 |
| 15 | 0.728011 | 2.15407 | 3.80789 | 2.15407 |

```
density.est(z = Newell.horiz, conf.level = 0.95)
```

```
##
##  Pollard's estimate of density using 4 sectors per point
##
## data:  Newell.horiz
## No. of sample pts: n = 15
## 95 percent confidence interval:
##   639.291 1062.625
## sample estimates:
## density per hectare
##             823.787
```

The transects and sample points are depicted in green. The nearest trees to each sample point are indicated with up or down triangles whose color alternates between red and blue corresponding to the transect. This may be used to identify whether the same tree is resampled from different points. The distances from the sample points to the nearest tree in each quarter can be listed and then analyzed using techniques discussed earlier. In this example, the 95% confidence interval for Pollard's method contains the actual density using the convex hull.

If the same corner points are used with vertical transects (`horizontal = FALSE`), the array of sample points is different. There are three points in each of five horizontal rows rather than five points in each of three horizontal rows as above. The density estimate using these vertical transects happens to be more accurate.

```
Newell.vert <- manual.survey(z = NewellSouth, t.coords = c(15, 8, 65, 8, 65, 26, 15, 26),
    points = 15, transects = 3, k = 1, horizontal = FALSE)
```

```
## Data: NewellSouth
##                      Area (ha) Density (per ha)
## Convex Hull              0.241          1056.00
## Bounding Rectangle       0.274           930.50
```

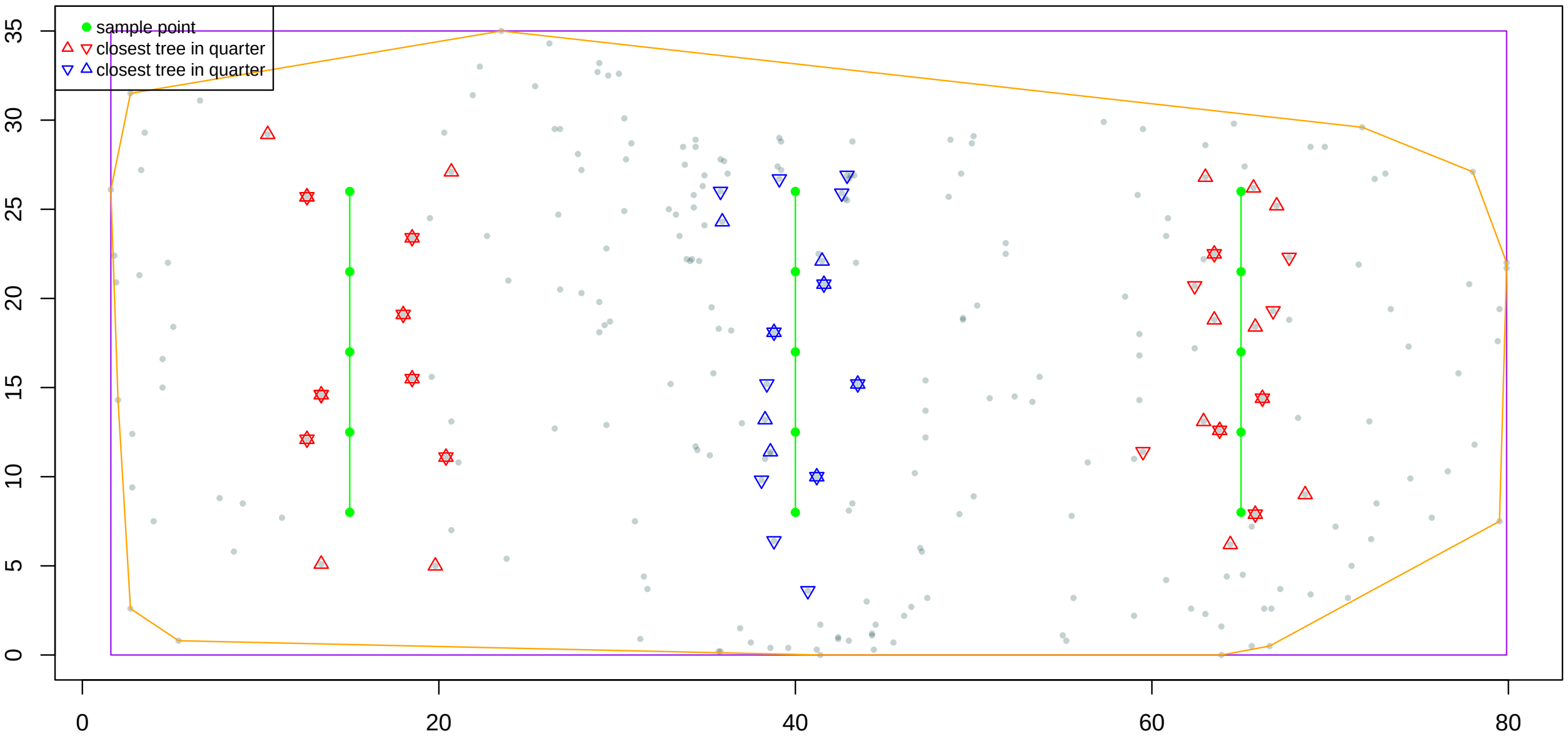

```
density.est(z = Newell.vert, conf.level = 0.95)
```

```
##
##  Pollard's estimate of density using 4 sectors per point
##
## data:  Newell.vert
## No. of sample pts: n = 15
## 95 percent confidence interval:
##   679.46 1129.39
## sample estimates:
## density per hectare
##             875.548
```

Morisita (1957) suggested measuring the distance to the third closest tree in each quarter (`k = 3`). For these data, this method gives quite a good estimate of the actual density of 1056 trees/ha.

```
Newell.third <- manual.survey(z = NewellSouth, t.coords = c(15, 8, 65, 8, 65, 26, 15, 26),
    points = 15, transects = 3, k = 3, display = FALSE)
angle.order.est(z = Newell.third, k = 3, method = "auto")
```

```
##
##  Morisita Practical Procedure for density estimate with 4 sectors at
##  each sample point using the third closest individuals. m1 > m2 (1045.75
##  > 968.82). Use m1.
##
## data:  Newell.third
## Number of sample points: n = 15
## sample estimates:
## density per hectare
##             1045.75
```

**EXAMPLE D.13** (Susquehanna Shale Hills Critical Zone Observatory)**.** The Susquehanna Shale Hills Critical Zone Observatory is an 8 hectare, first-order catchment in the Susquehanna River basin (see http://criticalzone.org/shale-hills/infrastructure/field-area/susquehanna-shale-hills-critical-zone-observatory/).

Among its publicly available resources is a data set of the locations of all trees in the basin (see Eissenstat and Kaye (2015)). The $(x, y)$-location data are available at http://math.hws.edu/pcqm/czo.csv. The data used below were modified (translated) by subtracting the minimum values of $x$ and $y$ from their respective coordinates to simplify the plot.

```
czo <- read.csv("http://math.hws.edu/pcqm/czoNormalized.csv")
czo.horiz <- manual.survey(z = czo, t.coords = c(75, 45, 400, 125, 420, 220, 95, 140),
    points = 24, transects = 4, k = 1)
```

```
## Data: czo
##                     Area (ha) Density (per ha)
## Convex Hull             8.507           243.80
## Bounding Rectangle     12.361           167.80
```

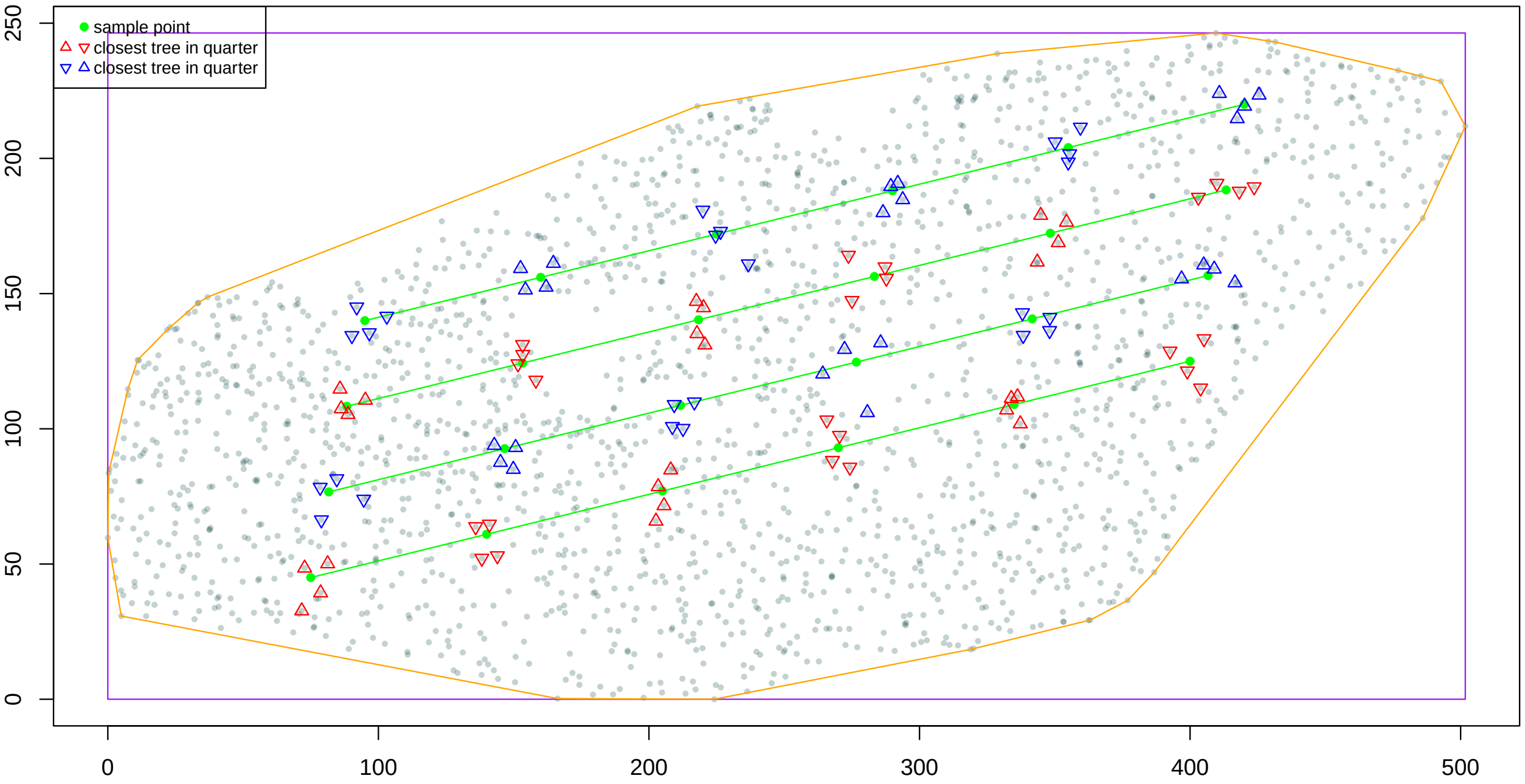

```
density.est(z = czo.horiz)
```

```
##
##  Pollard's estimate of density using 4 sectors per point
##
## data:  czo.horiz
## No. of sample pts: n = 24
## 95 percent confidence interval:
##  177.898 265.685
## sample estimates:
## density per hectare
##             217.339
```

Because of the shape of the plot, the bounding rectangle is not useful. The confidence interval for the density contains the actual density of 243.8 trees/ha determined by the convex hull.

The next analysis uses the same four corner points for the transects but with a vertical orientation and Morisita's recommendation for a third-order analysis. The resulting estimate is within 4% of the actual value.

```
czo.third.vert <- manual.survey(z = czo, t.coords = c(75, 45, 400, 125, 420, 220, 95, 140),
    points = 24, transects = 4, k = 3, horizontal = FALSE)
```

```
## Data: czo
##                     Area (ha) Density (per ha)
```

```
## Convex Hull              8.507           243.80
## Bounding Rectangle      12.361           167.80
```

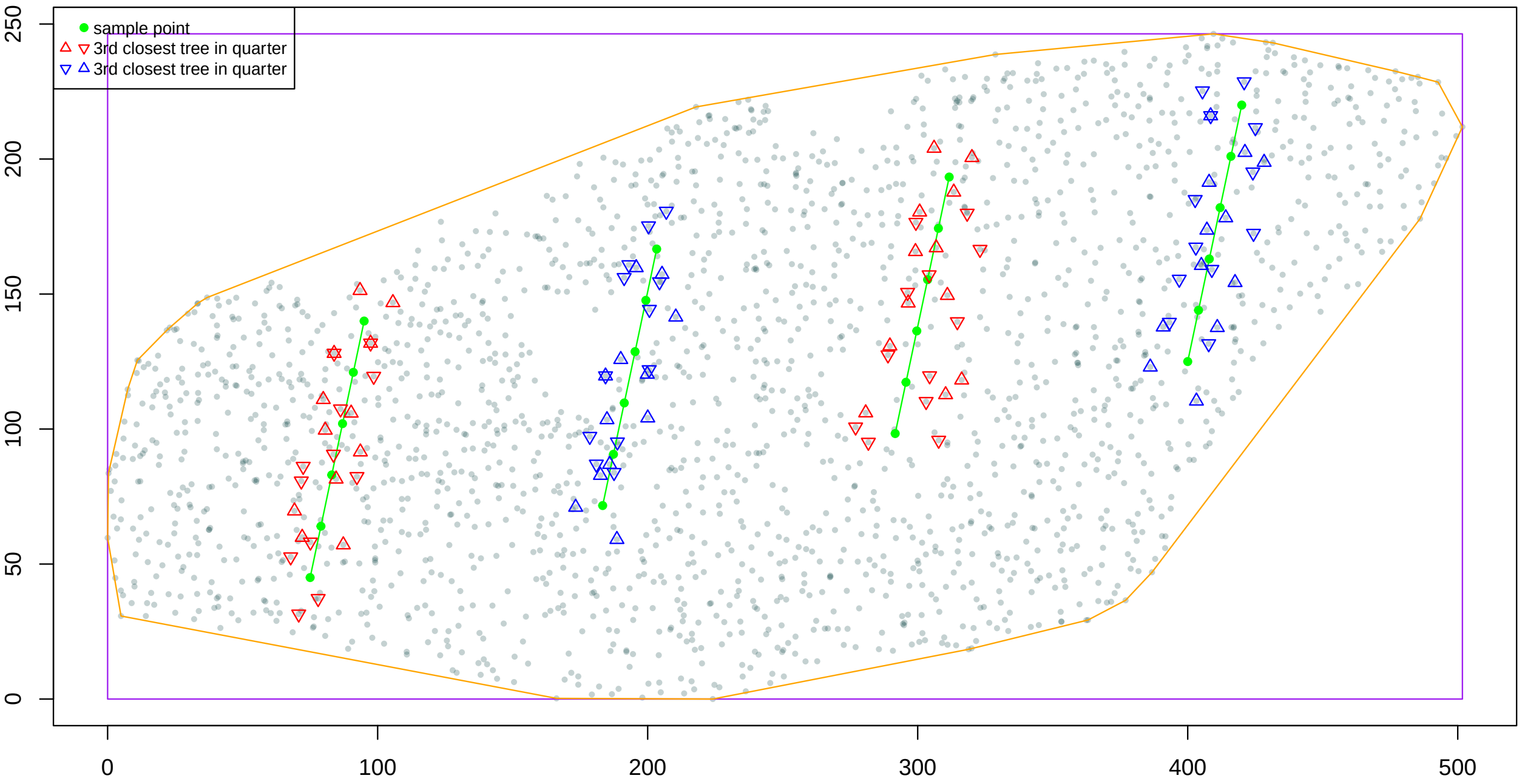

```
angle.order.est(z = czo.third.vert, k = 3, method = "auto")
```

```
##
##  Morisita Practical Procedure for density estimate with 4 sectors at
##  each sample point using the third closest individuals. m1 > m2 (234.65
##  > 232.19). Use m1.
##
## data:  czo.third.vert
## Number of sample points: n = 24
## sample estimates:
## density per hectare
##             234.651
```